\documentclass[a4paper, amsfonts, amssymb, amsmath, reprint, superscriptaddress, showkeys, nofootinbib, twoside, longbibliography]{revtex4-1}
\usepackage[T1]{fontenc}
\usepackage[polish,english]{babel}
\usepackage[utf8]{inputenc}
\usepackage[colorinlistoftodos, color=green!40, prependcaption]{todonotes}
\usepackage[colorlinks,breaklinks]{hyperref}
\usepackage{tabularx}
\usepackage{float}
\newcommand{\be}{\begin{equation}}
\newcommand{\ee}{\end{equation}}
\newcommand{\ben}{\begin{equation*}}
\newcommand{\een}{\end{equation*}}
\newcommand{\bea}{\begin{eqnarray}}
\newcommand{\eea}{\end{eqnarray}}
\usepackage{amsmath} 
\usepackage{graphicx}
\usepackage{bbm}

\newcommand{\abs}[1]{\ensuremath{\left| #1 \right|}}

\newcommand{\ket}[1]{\ensuremath{\left|\right.\!{#1}\!\left.\right\rangle}}

\newcommand{\bra}[1]{\ensuremath{\left\langle\right.\!{#1}\!\left.\right|}}
\newcommand{\braket}[2]{\ensuremath{\langle{#1}|{#2}\rangle}}
\newcommand{\ketbra}[2]{\ensuremath{|{#1}\rangle\langle{#2}|}}
\newcommand{\brakket}[3]{\ensuremath{\langle{#1}|{#2}|{#3}\rangle}}

\newcommand{\mm}[1]{\textcolor{black}{#1}}
\newcommand{\nh}[1]{\textcolor{black}{#1}}

\usepackage[framemethod=tikz]{mdframed}
\definecolor{mycolor}{rgb}{0.122, 0.435, 0.698}
\newmdenv[innerlinewidth=0.5pt, roundcorner=4pt,linecolor=mycolor,innerleftmargin=6pt,
innerrightmargin=6pt,innertopmargin=6pt,innerbottommargin=6pt]{mybox}

\usepackage{paracol}
\usepackage{tcolorbox}
\tcbset{colback=mycolor!5,colframe=mycolor,
fonttitle=\bfseries, float=htb}
\newtcolorbox[blend into=figures]{boxfigure}[3][]
{ float*=ht,width=\textwidth,lower separated=false, center upper,
title={#2},label= fig:#3,#1}
\newtcolorbox[blend into=figures]{smallboxfigure}[3][]
{float=ht,lower separated=false, blend before title=colon hang,
title={#2}, label= fig:#3 ,#1}
\newtcolorbox{smallbox}[3][]
{float=ht,lower separated=false, blend before title=colon hang,
title={#2}, label= fig:#3 ,#1}
\newtcolorbox[blend into=tables]{smallboxtable}[3][]
{float=ht,lower separated=false, blend before title=colon hang,
title={#2}, label= table:#3 ,#1}
\newtcolorbox[blend into=tables]{bigboxtable}[3][]
{float*=t,lower separated=false, blend before title=colon hang, width = 2\linewidth,
title={#2}, label= table:#3 ,#1}
\newcolumntype{Z}{|>{\centering\arraybackslash}X}

\hypersetup{
	bookmarksnumbered,
	pdfstartview={FitH},
	citecolor={darkgreen},
	linkcolor={darkred},
	urlcolor={darkblue},
	pdfpagemode={UseOutlines}}
\definecolor{darkgreen}{RGB}{50,190,50}
\definecolor{darkblue}{RGB}{0,0,190}
\definecolor{darkred}{RGB}{238,0,0}

\begin{document}
\title{Unscrambling Entanglement through a Complex Medium}

\author{Natalia Herrera Valencia}
\email[Correspondence email address: ]{nah2@hw.ac.uk}
    \affiliation{Institute of Photonics and Quantum Sciences (IPAQS), Heriot-Watt University, Edinburgh, UK}

\author{Suraj Goel}
    \affiliation{Institute of Photonics and Quantum Sciences (IPAQS), Heriot-Watt University, Edinburgh, UK}
    \affiliation{Indian Institute of Technology Delhi, New Delhi, India}

\author{Will McCutcheon}
    \affiliation{Institute of Photonics and Quantum Sciences (IPAQS), Heriot-Watt University, Edinburgh, UK}
    
\author{Hugo Defienne}
    \affiliation{School of Physics and Astronomy, University of Glasgow, Glasgow, UK}

\author{Mehul Malik}
    \email[Correspondence email address: ]{m.malik@hw.ac.uk}
    \affiliation{Institute of Photonics and Quantum Sciences (IPAQS), Heriot-Watt University, Edinburgh, UK}
    \affiliation{Institute for Quantum Optics and Quantum Information, Vienna, Austria}

\date{\today} 
\begin{abstract}
The transfer of quantum information through a noisy environment is a central challenge in the fields of quantum communication, imaging, and nanophotonics.
In particular, high-dimensional quantum states of light enable quantum networks with significantly higher information capacities and noise-robustness as compared with qubits. 
However, while qubit-entanglement has been distributed over large distances through free-space and fibre, the transport of high-dimensional entanglement is hindered by the complexity of the channel, which encompasses effects such as free-space turbulence or mode-mixing in multi-mode waveguides. 
Here we demonstrate the transport of six-dimensional spatial-mode entanglement through a two-metre long, commercial multi-mode fibre with 84.4\% fidelity.
We show how the entanglement can itself be used to measure the transmission matrix of the complex medium, allowing the recovery of quantum correlations that were initially lost. Using a unique property of entangled states, the medium is rendered transparent to entanglement by carefully ``scrambling'' the photon that did not enter it, rather than unscrambling the photon that did.
Our work overcomes a primary challenge in the fields of quantum communication and imaging, and opens a new pathway towards the control of complex scattering processes in the quantum regime.
\end{abstract}
\maketitle

\begin{figure}[b!]
\centering
\includegraphics[clip,trim=1.5cm 1.5cm 1.5cm 1.5cm,width=.9\linewidth]{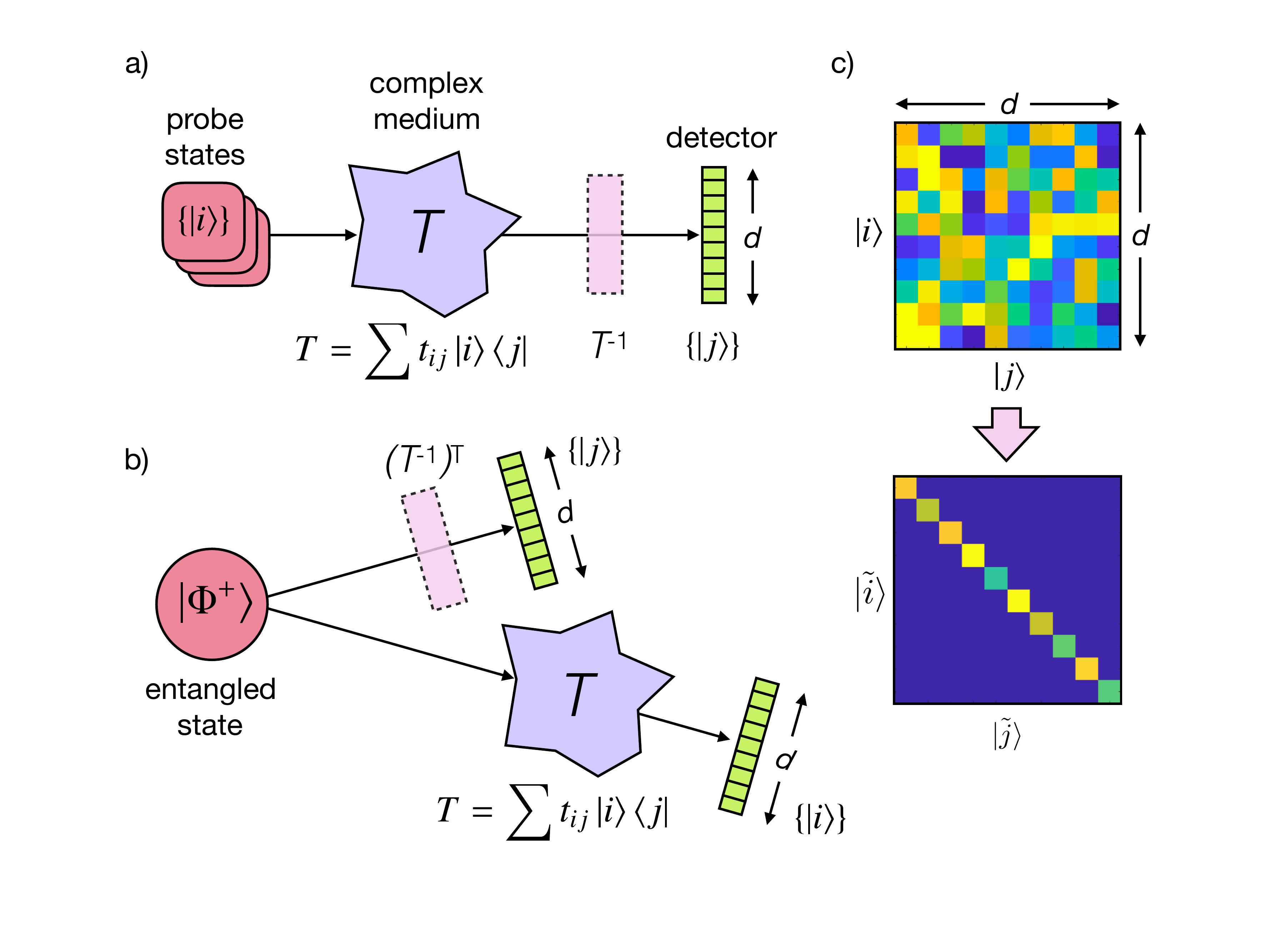}
\caption{Basic Principle. a) Classical methods for reconstructing a $d^2$-dimensional transmission matrix $T$ of a complex medium involve the measurement of a response function to a set of $d$ input probe states $\{\ket{i}\}$. b) Alternatively, the entire transmission matrix can be mapped onto a single maximally entangled state \ket{\Phi^+} of equal dimension.
 c) Information about $T$ can be used to reverse the effects of scattering by finding a set of propagation-invariant states $\{\ket{\tilde{i}}\}$, allowing one to send information through the medium. In the classical case, this involves applying an inverse operator such as $T^{-1}$ before or after the medium. To reverse the effects of scattering on entanglement, a suitable inverse operator can be applied on the photon that did not enter the medium at all.}
\label{basicidea}
\end{figure}

In recent years, the precise control of light propagation through disordered media has unlocked a range of new possibilities for biomedical imaging and optical telecommunications \cite{Rotter:2017gb}. The ability to turn a strongly scattering sample into a lens or send an image down an optical fibre no thicker than a human hair promises exciting new technologies such as non-invasive endoscopes and ultra-dense communication systems \cite{Papadopoulos:2013,Ploschner:2015fq, Turtaev:2018cw}. Key to achieving such control over light is the ability to measure the transmission matrix of a complex medium---a matrix of complex numbers that describes how the medium maps a set of input modes to a set of output modes \cite{Popoff:2010cj}. Enabled by the availability of highly tunable digital arrays, the transmission matrix is routinely measured today via the response of the medium to a set of probe states sent one at a time (Fig.~\ref{basicidea}a). Bringing this powerful technique to the domain of quantum information promises significant advances in quantum state transport and control. Recent experiments have demonstrated this potential by harnessing disorder for the manipulation of photon pairs \cite{Leedumrongwatthanakun:2019wt,Defienne:2016dk,Wolterink:2016bc} and transporting quantum correlations in a weak scattering regime \cite{Defienne:2018}.

Quantum entanglement plays a central role in the rapidly advancing field of quantum technologies \cite{Gisin:2007by}, enabling techniques such as quantum error correction and device-independent quantum communication \cite{Kelly:2015gi,Acin:2007db}. High-dimensional entangled states of light \cite{Krenn2017, Bavaresco:2018gw,Erhard:2018iua,Schneeloch:2019uw} offer vastly improved information capacities \cite{Mirhosseini:2015fy,Islam:2017hs} and greater resistance to noise \cite{Ecker:2019vx, Zhu:2019tb} over qubit-based quantum communication systems, and serve as a resource in quantum imaging protocols that allow one to image below the shot-noise level \cite{Brida:2010jw} or in an interaction-free manner \cite{Lemos2014}. High-dimensional entanglement can also tolerate large amounts of loss in loophole-free tests of nonlocality, holding immense potential for the realisation of device-independent quantum communication \cite{Vertesi:2010bq,Bavaresco:2017if}. 

A primary problem to be overcome in all of these applications is the preservation of the delicate quantum correlations found in entanglement after transmission through a channel. \mm{State-of-the-art demonstrations of entanglement transport include the distribution of qubit entanglement over 1200km of free-space and 300km of single-mode fibre \cite{Yin:2017ij,Inagaki:2013cx}, and dispersion-compensated qutrit entanglement over 1km of few-mode fibre \cite{Cao:2018uo}. However, the transport of entanglement through a complex scattering channel such as a multi-mode fibre or biological tissue remains a challenge. Such media involve the complex interplay of hundreds to millions of modes, and the effects of scattering must be overcome in a manner that preserves higher-order quantum coherence between all modes of interest. }


\begin{figure*}[t!]
\centering
\includegraphics[width=0.85\textwidth]{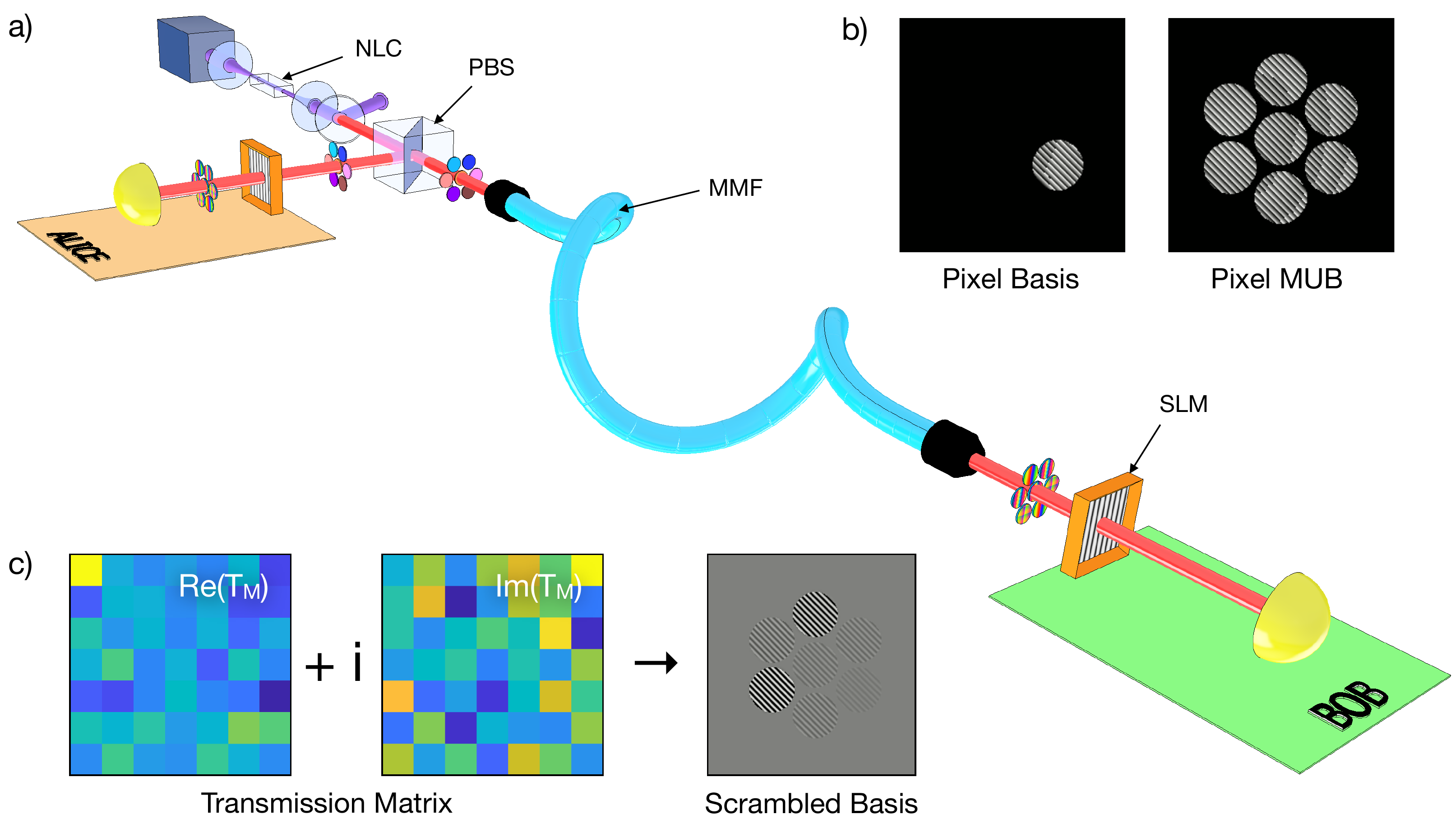}
\caption{Experimental setup and measurement holograms. a) A grating-stabilised UV laser is used to pump a nonlinear ppKTP crystal (NLC), producing pairs of photons entangled in their transverse position-momentum, which are separated by a polarising beam-splitter (PBS). One photon is sent into a 2-metre long commercial multi-mode fibre (MMF, Thorlabs GIF50E) and directed towards Bob, while its entangled twin is kept locally with Alice. Projective measurements of the resulting quantum state in any rotated Pixel basis are performed with spatial light modulators (SLM), single-mode fibres, and avalanche photo-diodes (not shown). b) Examples of holograms for measuring photonic states in the Pixel basis and its first mutually unbiased basis (Pixel MUB). As can be seen, the MUB hologram is made up of a coherent superposition of all seven macro pixels. c) Real and imaginary parts of the transmission matrix measured in the Pixel MUB ($T_M$) and an example of a ``scrambled'' basis hologram calculated from it. The scrambled basis hologram is displayed on Alice's SLM for unscrambling entanglement through the multi-mode fibre.}
\label{setup}
\end{figure*}

\mm{Here we demonstrate the transport of high-dimensional entanglement through a complex medium consisting of a short length of commercial multi-mode fibre}. In our experiment, the effects of scattering on entanglement are reversed by only manipulating the photon that did not enter the medium. To achieve this, we develop a new technique for measuring the transmission matrix of the fibre using the entangled state itself. In contrast to the classical technique of measuring a response function one state at a time, our method exploits the parallelism of entanglement by mapping the entire transmission matrix onto a single, high-dimensionally entangled state. \mm{This is known as the Choi-\foreignlanguage{polish}{Jamio\l{}kowski} isomorphism in quantum mechanics, which says that statements about a channel can be mapped onto statements about a state \cite{Jamiol:1972,DAriano:2001jl,Konrad:2007bs}. Previous experiments have used this isomorphism to characterise qubit processes \cite{Altepeter:2003gg} and quantum channels through classically non-separable states \cite{Ndagano:2017dz}. Here we apply it to complex, multi-mode scattering channels using high-dimensional entangled states, which are natural candidates for it.}

One particle of a maximally entangled two-particle state $\ket{\Phi^+}=\frac{1}{\sqrt{d}}\sum_i\ket{ii}$ is sent through the complex medium with transmission matrix $\hat{T} = \sum_{ij}t_{ji}\ket{j}\bra{i}$ (Fig.~\ref{basicidea}b). As a result, the two-particle entangled state undergoes a transformation into the pure state 
\be \ket{\Phi}_T=(\mathbb{I}\otimes \hat{T})\ket{\Phi^+}= \frac{1}{\sqrt{d}}\sum_{ij} t_{ji}\ket{ij},\label{eq:1}
\ee
\noindent which captures the entire knowledge of the medium's transmission matrix. Upon measuring $\ket{\Phi}_T$, one obtains the complex coefficients corresponding to $T$. The transmission matrix is thus obtained by characterising only one entangled state after it passes through the medium. As a result, this method requires $d$-times fewer settings for characterising a complex medium as compared with classical techniques, which require the preparation of $d$ separate probe states before the medium (Fig.~\ref{basicidea}a). Interestingly, entanglement is not strictly necessary for this method to work, however it does provide optimal results \cite{Altepeter:2003gg}.


\begin{figure}[b!]
\centering
\includegraphics[width=0.95\linewidth]{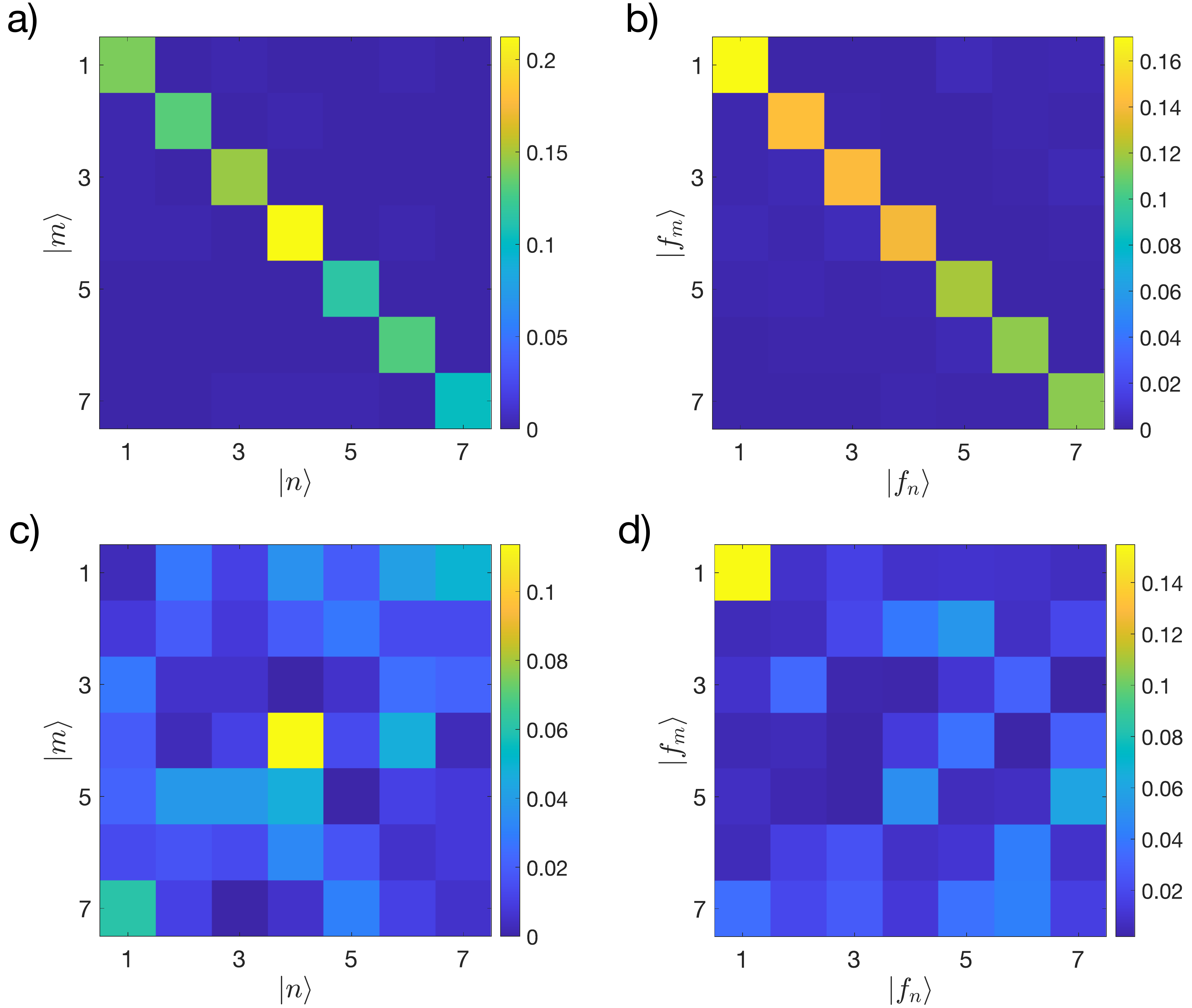}
\caption{Entanglement before and after a complex medium. In the absence of a multi-mode fibre, two-photon correlations measured in the a) Pixel basis and b) its first mutually unbiased basis (Pixel MUB) result in a fidelity to the 7-dimensional maximally entangled state of $F\geq94.1 \pm 1.2\%$, certifying the presence of 7-dimensional entanglement. The error on the fidelity bound is given by three times its standard deviation. c) Pixel and d) first Pixel MUB correlations after one photon from the entangled state is sent through a 2m-long multi-mode fibre, resulting in the loss of entanglement. Interestingly, even though the correlations are completely scrambled, the resulting data contains information about the transmission matrix in each basis ($\abs{T}^2$ and $\abs{T_M}^2)$ via state-channel duality (Eq.~\ref{eq:1}).}
\label{beforeafterMMF}
\end{figure}

Once a complex medium's transmission matrix is known, one can use it to reverse the effects of light scattering through the medium. This is done by either constructing a set of propagation-invariant states or eigenmodes \cite{Carpenter:2015, Ploschner:2015fq} obtained by diagonalising $T$ (Fig.~\ref{basicidea}c), or using the knowledge of $T$ to directly invert the scrambled light measured after the medium \cite{Popoff:2010dh}. 
Extending this methodology to the problem of unscrambling entanglement through a complex medium leads to an interesting revelation---one can invert the action of the complex medium by either unscrambling the photon that went through it, or by carefully ``scrambling'' the photon that did not (Fig.~\ref{basicidea}b). This results from a unique property of maximally entangled states where operations on the state can be equivalently expressed as being applied on either of its two parts \cite{Klyshko:1988eea}:
\bea
(\mathbb{I}\otimes T^{-1})\ket{\Phi}_T&=&(\mathbb{I}\otimes T^{-1}T)\ket{\Phi^+}= \ket{\Phi^+}\nonumber\\
&=&((T^{-1})^T\otimes T)\ket{\Phi^+}= \ket{\Phi^+}.
\eea

\noindent Thus, two-particle correlations lost due to one particle scattering through the medium are recovered by only manipulating the particle that did not enter the medium at all. This can also be understood as a consequence of the invariance of the state $\ket{\Phi^+}$ under transformations $ (U\otimes U^*)$ for any unitary operator $U$, \mm{which has been implicit in previous work on two-photon speckle \cite{Peeters:2010kx}} and used for the nonlocal cancellation of dispersion and weak scattering \cite{Franson:1992wo,Black:2019}.

While inverting the transmission matrix in this manner allows us to regain correlations in one basis, it does not guarantee that the state is entangled. To certify entanglement, one must be able to measure correlations in at least two complementary observables, or mutually unbiased bases (MUBs) of the state Hilbert space \cite{Wootters1989,Friis:2019hg}. Our measurement of $T$ relies on the assumption that the entangled state after the medium is pure ($\ket{\Phi}_T$ in Eq.~(\ref{eq:1}). Once $T$ is estimated, we must drop this assumption and use the measured $T$ to check how close to pure the transmitted state ($\rho_T$) actually is. In order to do so, we rotate our measured transmission matrix to any MUB, i.e. $T_M = M^* T M^T$, where $M$ is a complex matrix that performs the specified MUB transformation. Next, we use $T_M$ to construct a second ``scrambling'' operator on Alice, that should in principle allow us to recover correlations in the basis $M$ after transmission through the complex medium:
\be
( (T_M^{-1})^T M\otimes M^*)\ket{\Phi}_T= \ket{\Phi^+}.
\ee

Measurements in two or more mutually unbiased bases allow the use of a recently developed entanglement witness for certifying high-dimensional entanglement \cite{Bavaresco:2018gw}. Using this witness, we are able to lower bound the fidelity of the state to a given pure target state via measurements in two MUBs and certify entanglement dimensionality via a Schmidt number bound \cite{Malik:2015we}. Measurements in all MUBs allow us to calculate the exact fidelity to the target state, while also providing better noise performance.

\begin{figure*}[t!]
\centering
\includegraphics[width=0.9\linewidth]{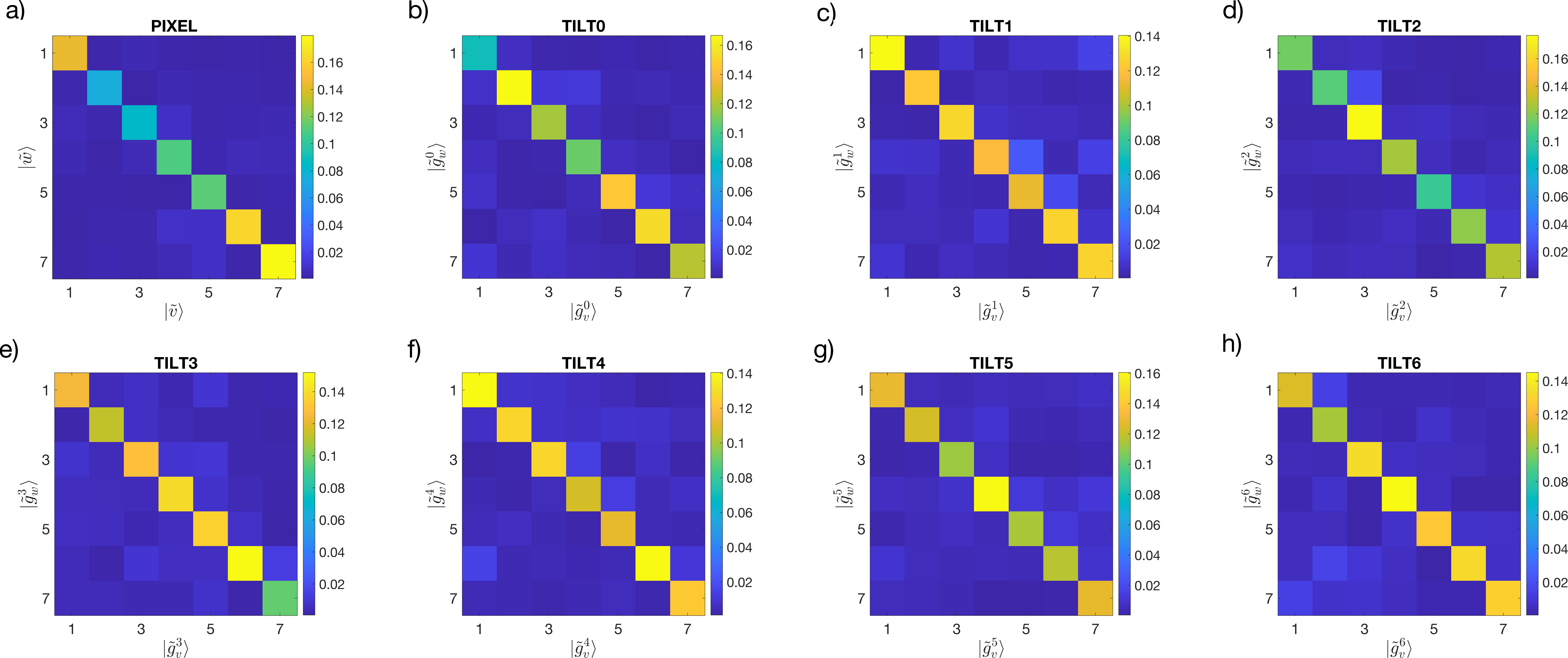}
\caption{High-dimensional entanglement unscrambled through a complex medium. Experimental data showing quantum correlations recovered in the a) Pixel and b-h) seven ``tilted'' Pixel bases after transmission through a multi-mode fibre. The Pixel basis measurements allow us to nominate a target state $\ket{\Psi}=\sum \lambda_i \ket{ii}$, and subsequently use its Schmidt coefficients $\lambda_i$ to construct MUB-like ``tilted'' bases following a recently developed \mm{entanglement certification} technique \cite{Bavaresco:2018gw}. Using these measurements, we can calculate the exact fidelity of the transported two-photon state ($\rho$) to the target state $(\ket{\Psi})$ to be $F(\rho_T,\Psi)=84.4\pm 1.8\%$, certifying the presence of six-dimensional entanglement. The error on the fidelity is given by three times its standard deviation.}
\label{tiltdata}
\end{figure*}

We perform an experimental test of our technique with states entangled in discretised transverse-position modes, also known as ``pixel'' entanglement \cite{Valencia2020}. As shown in Fig.~\ref{setup}a), photon pairs entangled in their transverse position-momentum are produced in a nonlinear crystal (NLC) via the process of spontaneous parametric downconversion (SPDC) and separated by a polarising beam-splitter (PBS). One photon is input into a commercial multi-mode fibre (MMF) and sent to Bob, while its entangled partner is kept with Alice. The MMF used in our experiment is a 2-metre graded-index fibre (Thorlabs GIF50E) that supports approximately 400 modes. Projective spatial-mode measurements of the resulting state at Alice and Bob are made with phase-only spatial light modulators (SLMs), single-mode fibres, and avalanche photo-diodes \cite{Bouchard:2018hr}. 

The Pixel basis used in our experiment is comprised of seven individual circular macro-pixels defined on the SLM. Fig.~\ref{setup}b) shows examples of diffractive holograms implemented on the SLMs for measuring states in the Pixel basis and first Pixel MUB. A particular mode $\ket{m}$ ($\ket{n}$) at Alice (Bob) is measured by displaying the corresponding macro-pixel hologram on their SLM. We can also define a family of orthonormal bases that are constructed from the Pixel basis using a transformation $\mathbf{M}=M_A \otimes M_B^*$, where $M$ takes us from the standard (Pixel) basis $\{\ket{m}\}_{m=0,\dots,d-1}$, to a new basis $\{\ket{f_k}\}^r_{k=0,\dots,d-1}$ given by:
\be 
\ket{f^r_k} = \frac{1}{\sqrt{d}}\sum_{m=0}^{d-1}\omega^{km+rm^2}\ket{m},
\label{eq:MUBstate}
\ee
with $k$ indexing a state in the new basis and $r=0,...,d-1$ indexing the basis itself. This construction follows the one introduced by Wootters and Fields \cite{Wootters1989}, which provides a set of bases mutually unbiased with respect to each other for prime dimensions. 

In the absence of a complex medium, two-photon correlations measured in the Pixel basis and first Pixel (Fourier) MUB respectively (Figs.~\ref{beforeafterMMF}a and b), certify a fidelity of $F\geq 94.1 \pm 1.2\%$ to a 7-dimensional maximally entangled state \cite{Bavaresco:2018gw}. Figs.~\ref{beforeafterMMF}c) and d) show measured correlations in these two bases after one photon from the state is sent through the multi-mode fibre, as shown in Fig.~\ref{setup}a). 
Correlations in both bases are completely lost, resulting in a trivial bound. Measurements in all eight MUBs after transmission through the fibre \nh{(See Extended Data Fig. 1)} result in an exact fidelity of $F(\rho,\Phi^{+})=5.4\pm 1.0\%$, which is lower than the bound of $F_2=1/7$ for two-dimensional entanglement (please see Methods for more details). Interestingly, while the measurements in Figs.~\ref{beforeafterMMF}c) and d) do not show any entanglement, they contain information about the absolute value of the fibre transmission matrix (Eq.~\ref{eq:1}), measured in the pixel basis and Pixel MUB ($\abs{T}^2$ and $\abs{T_M}^2)$ respectively.


A key challenge is retrieving the phase information of the transmission matrix elements using only intensity measurements. Here, we borrow a classical trick for doing so, where an internal reference mode is co-propagated through the medium and interfered with all the modes of interest \cite{Popoff:2010cj}. We use the space between our seven macro-pixels as a reference mode and vary its phase in four steps. The reference is characterised in a similar manner \nh{(please see Methods for more details)}. To maximise our count rates and minimise noise, we perform this procedure in the Pixel MUB instead of the Pixel basis. Via this phase-stepping holography, we are able to recover the complex values of the transmission matrix ($T_M$), and subsequently rotate to the other Pixel bases in order to certify entanglement. Fig.~\ref{setup}c) shows the real and imaginary parts of the measured transmission matrix, and an example of a ``scrambled'' basis hologram calculated from $(T_M^{-1})^T$ and displayed at Alice \cite{Bouchard:2018hr}. In this manner, entanglement is recovered by only manipulating the photon that did not enter the multi-mode fibre. 

Measurements showing the recovered state and its correlations in eight different bases are shown in Fig.~\ref{tiltdata}. Here, we use a recently developed adaptive witness that constructs MUB-like ``tilted'' bases to calculate the fidelity of the transmitted state $\rho_T$ to a non-maximally entangled target state $\ket{\Psi}=\sum_m\lambda_m\ket{mm}$ \cite{Bavaresco:2018gw}. The $r$-th tilted basis is defined by generalising the definition of a MUB (Eq.~\ref{eq:MUBstate}) in the following manner: 
\be 
\ket{\tilde{f}^r_k} = \frac{1}{\sum_n\lambda_n}\sum_{m=0}^{d-1}\omega^{km+rm^2}\sqrt{\lambda_m}\ket{m},
\label{eq:TMUBstate}
\ee
where $\lambda_m$ refer to the Schmidt coefficients of the non-maximally entangled target state. Notice that for $\sum_m \lambda_m^2=1$, the basis vectors $\ket{\tilde{f}^r_k}$ are normalised but not necessarily orthogonal. This construction satisfies the condition $|\braket{m}{\tilde{f}^r_k}| = \lambda_{m}\lambda_{k} \forall m,k$ with the standard basis $\{\ket{m}\}_{m=0,\dots,d-1}$. \mm{Our recovered state after the multi-mode fibre (Fig.~\ref{tiltdata}a) is non-maximally entangled owing to the non-unitarity of the channel, while still retaining its purity. The tilted-basis witness and an appropriately chosen target state are thus quite suitable for certifying entanglement in this scenario (please see Methods for more details).} Using the measurements shown in Fig.~\ref{tiltdata}, we are able to calculate an exact fidelity to the target state estimated from Fig.~\ref{tiltdata}a) of $F(\rho_T,\Psi)=84.4\pm 1.8\%$, certifying the presence of six-dimensional entanglement. The uncertainty in the fidelity is calculated assuming Poisson counting statistics and propagating the error via a Monte-Carlo simulation of the experiment.

In order to achieve the transport of high-dimensional entanglement through a multi-mode fibre, our experiment employed the use of single-outcome spatial-mode measurements that were scanned through the basis of interest. Consequently, this limits the speed with which the complex medium or entangled state can be characterised. Furthermore, our multi-mode fibre was quite short, limiting the effects of spatial-mode dispersion. Rapid progress in the development of quantum technologies, \mm{such as generalised mode transformers \cite{Fontaine:2019ja}, single-photon detector arrays \cite{Wollman:2019wa}, and spatio-temporal wavefront shaping approaches \cite{Mounaix:2016}}, should allow our technique to be used for entanglement transport through highly dynamic scattering samples, such as living biological tissue or km-long multi-mode fibre \cite{Rotter:2017gb}. Furthermore, our work can be generalised to the case of both photons traveling through two independent channels, with only one particle being manipulated (please see Supplementary material for more details). Such an ability could be useful in quantum network scenarios \cite{Epping:2017dx} or for non-invasive biological imaging \cite{Kang:2015}, where access to all parts of the complex system may be limited. Our results thus have immediate ramifications for the fields of quantum communication and imaging \cite{Diamanti:2016bu,Moreau:2019ca}, where the transport and control of quantum states of light through complex media remains a pressing challenge.
\newpage

\textit{Acknowledgements.} We thank M.~Huber, N.~Friis, D.~Phillips, S.~Leedumrongwatthanakun, and A.~Fedrizzi for helpful discussions.
This work was made possible by financial support from the QuantERA ERA-NET Co-fund (FWF Project I3773-N36) and the UK Engineering and Physical Sciences Research Council (EPSRC) (EP/P024114/1). H.D. acknowledges funding from the European Commission via a Marie Curie project (840958).

\textit{Author Contributions.} M.M.~conceived the research and supervised the project. M.M.~and H.D.~designed the experiment. N.H.V.~and S.G.~performed the experiment. All authors developed theoretical methods, analysed the data, and contributed to writing the manuscript.

\textit{Competing interests}
The authors declare no competing interests.


\newpage
\noindent\textbf{METHODS}
\subsection*{Mapping the multi-mode fibre channel onto an entangled state}

A complex medium such as a multi-mode fibre (MMF) acts as a scattering channel. If one particle of a bipartite high-dimensional entangled state enters the channel, its quantum correlations are affected by mode-mixing and cross-talk. To reverse this effect, we will show how the information of the channel is mapped onto the output state, allowing us to determine the transmission matrix that characterises the scattering process in the fibre.

Let us consider a general bipartite state
\be
\ket{\psi_{\text{i}}}=\sum_{ij}C_{ij} \hat a^\dagger_{i} \hat b^\dagger_{j} \ket{\text{vac}} = \sum_{ij}C_{ij} \ket{i}_A\ket{j}_B,
\label{eq:genstate}
\ee
where $i$ and $j$ label the spatial modes of the biphoton state shared by two parties, Alice and Bob. The state lives in a Hilbert Space of dimension $d=7$, which is smaller than the amount of modes supported by the MMF channel. Let us thus divide the state space into a logical subspace, corresponding to the modes that we measure, and environmental/loss modes, which are not considered in this process. Adding two extra indices $n$ and $m$ to our initial state to indicate whether the mode is in the logical or environmental subspace, we can write
\bea
\ket{\psi_{\text{i}}}&=&\sum_{injm}C_{injm} \hat a^\dagger_{in} \hat b^\dagger_{jm} \ket{\text{vac}} \nonumber \\
&=& \sum_{injm}C_{injm} \ket{in}_A\ket{jm}_B,
\label{eq:init}
\eea
where a given $i,j$ mode is in the logical subspace if $m,n =1$, or in the environmental subspace otherwise.

In our system, Bob's modes go through the MMF and undergo the unitary transformation
\be
\hat{U}_{\text{MMF}} = \sum_{klrs}U_{klrs}\ketbra{kl}{rs}.
\ee
The matrix $U$ is given by elements $U_{klrs}$ that describe how the modes $\ket{rs}$ scatters to the modes $\ket{kl}$. The action of the scattering on our initial state is given by
\be
\hat b^\dagger_{jm} \xrightarrow{U} \hat U \hat b^\dagger_{jm} \hat U^\dagger= \sum_{kl} U_{kljm} \hat b^\dagger_{kl}.
\ee
After the MMF, we perform measurements in the $d$-dimensional logical subspace. This measurement and postselection, which results in a state conditioned on both photons being in the logical subspace ($n,m=1$), can be described using the following projector:
\be
\begin{split}
\hat \Pi &= \sum_{pq} \hat a^\dagger_{p1} \hat b^\dagger_{q1} \ketbra{\text{vac}}{\text{vac}} \hat b_{q1} \hat a_{p1}\\
&=\sum_{pq} \ket{p1}_A {\ket{q1}_B} _B\bra{p1} _A\bra{q1}.
\end{split}
\label{eq:projector}
\ee
This occurs in general with non-unit success probability, resulting in the sub-normalised state after the MMF given by:
\be
\begin{split}
\ket{\psi}_{\text{MMF}}&:=\hat \Pi (\hat{I}_A \otimes \hat U_{\text{MMF}}) \ket {\psi_{\text{i}}}\\ 
&= \sum_{ijmk} U_{k1jm} C_{i1jm}\ket{i1}_A\ket{k1}_B\\
&:= \sum_{ik}t_{ki} \ket{i}_A \ket{k}_B,
\end{split}
\label{eq:PsiMMF}
\ee
where we define the coefficients characterizing the state after the fibre as $t_{ki} := \sum_{jm}U_{k1jm}C_{i1jm}$. Notice that these coefficients encode the information of both the MMF and the state. 

As shown in the main text, the state we produce through spontaneous parametric down conversion is very close to the maximally entangled state:
\be
\ket{\Psi^{+}} =\frac{1}{\sqrt{d}}\sum_{i} \ket{i}_A\ket{i}_B.
\label{eq:MaxEnt}
\ee
We find in this case that the action of the unitary channel representing the MMF, followed by postselection onto states with photons detected in the logical subspace, is characterised by the operator $\hat T$ with elements $t_{ki}=U_{k1i1}$:
\be
\hat{T}_{\text{MMF}} = \sum_{ki} U_{k1i1}\ketbra{k}{i},
\ee
whose coefficients describe how the logical mode $i$ scatters into the logical mode $k$. This is not a unitary operation (since it is only a sub-matrix of the full unitary transformation), and is not trace preserving (as the postselection happens in general with non-unit success).

In this sense, even though our MMF is a unitary channel on \emph{all} the modes that the fibre can support, since we are interested only in the logical modes, the operator $\hat{T}$ acts as the relevant non-unitary transmission matrix of the fibre. Since this operator is non-unitary it has the effect of changing the entanglement in the state, despite only acting locally on Bob's modes.  

If our initial state is maximally entangled as in Eq.~\ref{eq:MaxEnt}, and we consider this operator that represents the scattering effect of the fibre on all the modes in Bob's space (logical and environmental), the state after one of the photons of the entangled pair goes through the fibre is given by:
\be
\ket{\psi_{\text{MMF}}} =(\hat{I}\otimes \hat{T})\ket{\psi^{+}} = \sum_{ik} U_{k1i1}\ket{i}_A\ket{k}_B
\label{eq:Choi}
\ee 
This is precisely the Choi-\foreignlanguage{polish}{Jamio\l{}kowski} isomorphism, where the channel representing the fibre is imprinted onto the initial entangled state.

We note that this subnormalised state is \emph{pure} despite having undergone a nonunitary transformation. In the absence of postselection, one could instead model the trace-preserving channel acting only on the logical modes by including the vacuum state in the output Hilbert space. In the Kraus representation, this trace-preserving channel is given by the Kraus operators
\be
\begin{split}
A_0 &= U_{k1r1}\ket{k}\bra{r}\\
A_m &= U_{m2r1}\ket{\text{vac}}\bra{r} \\
\text{so that }\,\, \sum_{\forall m} A_m^\dagger A_m &= \hat{I},
\end{split}
\label{eq:KrausTP}
\ee
which is clearly not a pure channel if any $U_{m2r1}\neq0$, i.e. if there is any loss, as the output state is a mixture with the vacuum. However, after postselection on the existence of a photon, only the term originating from $A_0$ survives, resulting in a postselected subnormalised pure state. 

It is clear from Eqs.~\ref{eq:PsiMMF} and \ref{eq:Choi} that considering the initial state to be a general entangled state, or the maximally entangled state, leads to analogous resulting states after the fibre $\ket{\psi_{\text{MMF}}}$. Even more, if one attributes the effect of coefficients $C_{ij}$ to the channel $U$, the two cases are the same. Because of this, we consider our initial state to be $\ket{\Psi^+}$.

\subsection*{Experimental details}

A continuous-wave grating-stabilised laser (Toptica DL Pro HP) at 405~nm pumps a periodically poled Pottasium Titanyl Phosphate (ppKTP) crystal (1~mm~$\times$~2~mm~$\times$~5~mm) at 75~mW to generate a pair of orthogonally polarised photons at 810~nm entangled in their position-momentum degree of freedom (DOF) through the process of Type-II spontaneous parametric down conversion (SPDC). Phase-matching conditions are achieved via temperature-tuning the crystal in a custom-built oven that keeps it at 30$^\circ$C. A 5:1 telescope system of lenses is used to shape the pump beam and focus it on the crystal with a $1/e^2$ beam diameter of 400~$\mu$m. A dichroic mirror (DM) removes the pump after the crystal, while the pair of produced photons are separated by a polarising beam splitter (PBS). The reflected photon (corresponding to Alice) has its polarisation rotated from vertical to horizontal with a half-wave plate (HWP) and made incident on a phase-only SLM (Hamamatsu X10468-02) that is placed in the Fourier plane of the crystal using a 250~mm lens. The transmitted photon (corresponding to Bob) is shaped with lenses in order to effectively couple modes in our 7-dimensional Hilbert space of interest into a 2~m graded-index multimode fibre (Thorlabs M116L02) that has a core diameter of 50~$\mu$m and supports around 400 spatial modes at 810~nm. After going through the fibre, the photon is launched onto another phase-only SLM.

Measurements in the position-momentum DOF of the photons are made with computer generated holograms (CGH) displayed by the parallel-aligned liquid-crystal-on-silicon (LCOS) layer of the SLMs, which has an effective area size of 15.8 $\times$ 12 mm, pixel pitch of 20~$\mu$m, resolution of 792 $\times$ 600, a reflection efficiency of approximately 90~\%, and a diffraction efficiency of approximately 75~\%. The CGH in combination with a single-mode fiber (SMF) allows us to projectively measure whether the incident photons are in a particular spatial mode in any given basis. The accuracy of the projections performed with the combo of SLM and SMF is ensured through the use of an intensity flattening telescope \cite{Bouchard:2018hr}. This technique removes the Gaussian component introduced by the use of SMF by afocally decreasing the size of the mode propagating from the SLM to the objective lens, thus recovering the orthogonality relation between spatial modes of a given basis. The SMFs guide the filtered photons to single-photon avalanche detectors (Excelitas SPCM-AQRH-14-FC) with an efficiency of 60~\% at 810~nm. The detectors are connected to a coincidence counting (CC) logic (UQD Logic16) that records time-coincident events within a window of 0.2~ns.

\subsection*{Measurement of the transmission matrix}

As discussed in the main text, when one of the photons of a high-dimensional entangled bi-photon state is sent through the MMF, the effect of this scattering channel is encoded onto the complex coefficients $t_{ij}$ characterizing the output state $\ket{\psi_{\text{MMF}}}$. These coefficients form a matrix $T$ that we consider as the effective or relevant transmission matrix for the modes of the $d-$dimensional subspace we measure in.

We use a phase-stepping technique in order to determine $T$. For doing so, let us define a mode number 0 as our internal reference mode. The reference mode in our experiment was taken to be a ``background'' mode, which is defined by all the SLM pixels that lie between the macro-pixel modes. The phase-stepping process is implemented by displaying a relative phase $\theta$ between the basis and the reference modes. In this manner, when Alice scans the mode $m$, the reference mode is displayed simultaneously with a controlled phase difference. On Bob's side, we will simply display the mode $n$ \nh{(See Extended Data Fig. 2)}. 

The state we are projecting on with each measurement of one element of the correlation matrix is given by:
\be
\ket{\chi_{mn}} = (e^{i\theta}\ket{0}+\ket{m})_A \otimes \ket{n}_B,
\ee
where $m,n = 1,...,d$ represent basis elements of the Pixel basis. From these measurements, we can construct a matrix $R$ with coefficients of the form:
\be
R_{mn}(\theta)= | \bra{\chi_{mn}} \Psi_{MMF} \rangle|^2 = \left | t_{n0} e^{-i\theta} + t_{nm} \right |^2 
\label{eq:Rmn}
\ee
Setting $\theta$ in steps of $\pi/2$ going from 0 to $3\pi/2$, we obtain:
\bea
S_{mn} &:=& \frac{1}{4}\left [R_{mn}^{0} - R_{mn}^{\pi} +i(R_{mn}^{\pi/2} - R_{mn}^{3\pi/2}) \right] \\ \nonumber
&=& t_{n0} t^*_{nm}= E_n t^*_{nm} 
\eea
The resulting matrix $S$ is not exactly equal to the transmission matrix $T$, but related to it as follows:
\bea
\mathbf{S}=\left(
\begin{matrix}
t^*_{11} & \dots & t^*_{d1}\\
\vdots & \ddots & \vdots \\
t^*_{1d} & \dots & t^*_{dd}
\end{matrix}
\right)
\left(
\begin{matrix}
t_{10} & \dots & 0\\
\vdots & t_{j0} & \vdots \\
0 & \dots & t_{d0}
\end{matrix}
\right) = \mathbf{T^\dagger E},
\label{Smatrix}
\eea 

\noindent where $E$ is a diagonal matrix related to the mixing of the reference with the basis modes after going through the MMF. Determining $E$ is crucial for fully recovering the correlations of our entangled state. In order to do so, we again use the phase-stepping technique where we now only display the reference mode on Alice, while Bob simultaneously displays a basis and reference mode \nh{(See Extended Data Fig. 3)}:
\be
\ket{\chi}_m =\ket{0}_A \otimes (e^{i\theta}\ket{0}+\ket{m})_B 
\ee

The results of these measurements are given by:
\be 
R_m^\theta = |\braket{\chi_m}{\Psi_{MMF}}|^2 = 
|t_{00}e^{i\theta}+ t_{m0}|^2 
\ee

Performing the measurement for different relative phases, we recover the following terms
\bea 
\tilde{E}_{m} &=&\frac{1}{4}\left[(R_m^0 - R_m^\pi+i(R_m^{\pi/2}-R_m^{3\pi/2})\right] \nonumber \\
&=& t_{00}t_{mo}^* = t_{00}E_m^*.
\label{eq:Em}
\eea

Building a diagonal matrix from the terms $\tilde{E}_m$, we have:
\be 
\mathbf{\tilde{E}} = t_{00}
\left( \begin{matrix}
E_{1}^* & \dots & 0\\
\vdots & E_{j}^* & \vdots \\
0 & \dots & E_{d}^*
\end{matrix} \right) = t_{00} \mathbf{E^*} 
\ee 

The matrix $\mathbf{\tilde{E}}$ is equal to the conjugate of matrix $\mathbf{E}$ with a factor of $t_{00}$ that we neglect because it only adds a global amplitude and phase. After determining both $S$ and $E$, we can calculate the $T$ matrix characterising the effect of the MMF as follows:
\be 
\mathbf{T = (SE^{-1})^\dagger }
\label{TMeas}
\ee 

It is important to notice that because we are using localised transverse spatial (or momentum) modes, the single-outcome measurements we have been describing are limited by their projection onto the collection mode, leading to lower counts when using the standard basis than when using any of the MUBs~\cite{Valencia2020}. To have a faster and less noisy measurement of the transmission matrix, we thus use the first MUB basis (corresponding to $r=0$ in the Wootters-Fields construction of Eq.~\ref{eq:MUBstate}) for the phase-stepping process. 

It can be easily proven that using a basis different than the standard when following the steps described above simply results in a rotation of the transmission matrix. In our case, the obtained transmission matrix corresponding to measurements in the first MUB is given by:
\be 
T_{M_0} = M_0^* T M_0^T,
\label{eq:Tm}
\ee
where $T$ is the transmission matrix that one would determine had the measurement been made in the standard basis and $M_0$ is the MUB transformation matrix.

\subsection*{Unscrambling entanglement}

The knowledge of the transmission matrix of the fibre allows us to construct a new set of measurement bases that invert the mixing process of the modes of the entangled state, thus recovering or ``unscrambling'' the quantum correlations present in $\ket{\psi_{\text{i}}}$. First we will consider the case in which the transmission matrix was determined in the standard basis, and then we will generalize for any rotation of $T$.

Interestingly, this new set of bases can be used either by Alice or Bob for recovering the state. Instead of unscrambling the mode-mixing on Bob's side, we choose to ``scramble'' the modes on Alice's side to show that even though Bob's photon is the one going through the MMF, by performing the right measurements on Alice, the correlations of the entangled bi-photon state are recovered. 

The bi-photon state after propagation through the fibre is described by the state:
\be
\ket{\psi_{\text{MMF}}} = (\hat{I}_A \otimes \hat{T}){\ket{\Psi^+}}.
\ee 
Reversing the action of $T$ on the entangled state can be done simply by applying the inverse operation on the photon going through the fibre:
\be
(\hat{I}_A \otimes \hat{T}^{-1})\ket{\psi_{\text{MMF}}} = {\ket{\Psi^+}}.
\ee 
Of course since $\hat T\preceq \hat I$ (the singular values of the matrix $T$ are all less than 1) the inverse does not constitute a physical channel, and must be appropriately normalised. We defer this discussion to the supplementary material. Since we want to only manipulate the state on Alice's side, we consider the following property:
\be 
(\hat{A} \otimes \hat{B})\ket{\Psi^{+}} = (\hat{B}^T\hat{A} \otimes \hat{I})\ket{\Psi^{+}} = (\hat{I} \otimes \hat{B}\hat{A}^T)\ket{\Psi^{+}}.
\label{MEProperty}
\ee
Hence, for inverting the action of the MMF, one can use Eq.~\ref{TMeas} to generate an operator on Alice's side:
\be
W_A= (T^{-1})^T = (((SE^{-1})^\dagger)^{-1})^T.
\label{eq:Woperator}
\ee 
In this manner, we can define an operator $\mathbf{W} = \hat{W}_A\otimes \hat{I}_B$ that converts the output state of the MMF to the initial maximally entangled state:
\bea
\ket{\psi}_{\text{u}}&=&\mathbf{W}\ket{\psi_{\text{MMF}}} \nonumber \\
&=&(\hat{W}_A\otimes \hat{I}_B)\ket{\psi_{\text{MMF}}} \nonumber \\
&=& ((\hat{T}^{-1})^T\otimes \hat{I}_B)\ket{\psi_{\text{MMF}}} \nonumber \\
&=& (\hat{I}_A\otimes \hat{T}^{-1})\ket{\psi_{\text{MMF}}} \nonumber \\
&=& \ket{\Psi^{+}}.
\eea
While on Bob's side we use the standard basis, the rows of the operator $\hat{W_A}$ constitute a new basis to be used by Alice. Strong correlations should appear in the cross-talk matrix obtained from measuring two-photon coincidences with these bases.

The second step for certifying entanglement is to recover correlations in a mutually unbiased or a tilted basis. Using $\hat{W}_A$ and either of the transformations $M_r$ ($r=0,...,d-1)$ or $\tilde{M}_r$ ($r=0,...,d-1)$, which define the states given by Eq.~\ref{eq:MUBstate} and Eq.~\ref{eq:TMUBstate} respectively, we can define a new operator to unscramble the correlations in the $r$-th mutually unbiased basis:
\bea
\mathbf{V} &=& \mathbf{M_rW} \nonumber \\
&=& (\hat{M}_r\otimes \hat{M}_r^*)(\hat{W_A} \otimes \hat{I}_B) \nonumber \\
&=& (\hat{M}_r\hat{W}_A \otimes \hat{M}_r^*) \nonumber \\
&=& (\hat{V}_A \otimes \hat{M}_r^*).
\label{eq:Voperator}
\eea
If we use $\tilde{M}_r$ instead of $M_r$, we unscramble the correlations in the $r$-th tilted basis. Using the rows of $\hat{V}_A$ as a new measurement basis for Alice leads to recovered correlations of the initial state in a MUB or tilted basis. Measuring in the new bases defined by $\hat{W}_A$ and $\hat{V}_A$ at Alice, and the standard and mutually unbiased basis at Bob, should result in strong correlations in two bases that allow us to certify high-dimensional entanglement. 

If one uses a basis different from the standard basis for the phase-stepping process, the transmission matrix recovered is a rotated version of the transmission matrix in the standard basis $T$. In this case, the construction of the new bases used for recovering quantum correlations must also include a rotation back to the basis that is being used on Bob.

The transmission matrix in the standard basis can be expressed in terms of $T_M$ as $T=M^TT_MM^*$, and thus, the operator $\mathbf{W}$ can be written as:
\be
\mathbf{W} = \left(((M^T T_M M^*)^{-1})^T  \otimes \hat{I}_B\right)
=\left(M^{\dagger}(T_M^{-1})^TM  \otimes \hat{I}_B\right) 
\ee
Notice that performing the measurements with $\mathbf{W}$ for getting the correlations in the standard basis would require one to use standard basis modes on Bob's side, which, as we mentioned before, leads to lower counts. 

As an alternative, one can consider an operator where the rotation of the transmission matrix to the standard basis is made partially on Alice, while in Bob one uses the MUB:
\be
\mathbf{W_M} = \left((T_M^{-1})^TM  \otimes M^*\right) 
\label{eq:Wmoperator}
\ee
In this case, both SLMs display multiple pixels and we get a higher level of counts for this measurement. Applying this unscrambling operator to the MMF state equivalently leads to the maximally entangled state:
\be
\begin{split}
\ket{\psi}_{\text{u}}&=\mathbf{W_M}\ket{\psi_{\text{MMF}}} \\
&= \left((T_M^{-1})^TM  \otimes M^*\right) \ket{\psi_{\text{MMF}}} \\
&= \left((T_M^{-1})^TM  \otimes M^*\right)(I_A \otimes \hat{T}) \ket{\Psi^+} \\
&= \left((T_M^{-1})^TM  \otimes M^*T\right) \ket{\Psi^+} \\
&= \left(I_A  \otimes M^* T M^T (T_M^{-1})\right) \ket{\Psi^+} \\
&=(I_A \otimes T_M(T_M^{-1}))\ket{\Psi^+} \\
&=\ket{\Psi^+}. \\
\end{split}
\ee
Analogous to $\mathbf{W}$, using the operator $\mathbf{W_M}$ results in an unscrambled state that is maximally entangled. This means that the unscrambled state in the standard basis can also be recovered by partially rotating the transmission matrix at Alice, and completing the rotation by measuring in a MUB at Bob. 
The normalised two-photon coincidences obtained in this case are given by:
\bea
N_{wv} &=& |\braket{wv}{\psi_{\text{MMF}}}|^2 \nonumber \\
&=& |\brakket{mn}{\mathbf{W_M}}{\psi_{\text{MMF}}}|^2 \nonumber \\
&=&  |\braket{mn}{\psi_u}|^2,
\label{eq:Nwv}
\eea
where the new scrambled basis elements $\ket{w}$ (For Alice) and $\ket{v}$ (For Bob) are calculated by applying the operator $\mathbf{W_M}$ to the standard basis elements $\ket{m}$ and $\ket{n}$. The cross-talk matrix resulting from these measurements should show strong correlations that correspond to the ones in the standard basis.

For the second measurement, let us consider that we measured the transmission matrix in the basis indexed by $r=0$. In this case, the operator $\mathbf{V}$ that recovers correlations in a basis $r$ can be written in terms of $T_{M_0}$ as:
\be 
\mathbf{V_{M}}=\mathbf{M_r W_M}= (M_r\otimes M_r^*)\left((T_{M_0}^{-1})^T M_0  \otimes M_0^*\right), 
\label{eq:Vmoperator}
\ee
where we simply add to the first measurement the $\mathbf{M_r}$ transformation corresponding to the basis in which we want to recover the correlations. The normalised two-photon coincidences  are given in this case by:
\bea
\tilde{N}_{wv} &=& |\braket{g_wg_v}{\psi_{\text{MMF}}}|^2 \nonumber \\
&=& |\brakket{mn}{\mathbf{V_{M}}}{\psi_{\text{MMF}}}|^2 \nonumber \\
&=& |\brakket{f_mf_n}{\mathbf{W_{M}}}{\psi_{\text{MMF}}}|^2 \nonumber \\
&=& |\braket{f_mf_n}{\psi_u}|^2.
\label{eq:tNwv}
\eea
Notice that the basis elements used at Alice ($\ket{g_w}$) and Bob ($\ket{g_v}$) are calculated by applying the operator $\mathbf{V_M}$ to the elements of the MUB basis $\ket{f_m}$ and $\ket{f_n}$. If one uses the tilted basis, i.e., $\mathbf{\tilde{M}_r}$ instead of $\mathbf{M_r}$, the new basis elements are $\ket{\tilde{g}_w}$) and $\ket{\tilde{g}_v}$, which are calculated by applying the operator $\mathbf{V_M}$ to the elements of the tilted basis $\ket{\tilde{f}_m}$ and $\ket{\tilde{f}_n}$.

\subsection*{Data availability} 
All data that support the plots within this paper and other findings of this study are available from the corresponding authors upon reasonable request.

\clearpage
\onecolumngrid
\appendix
\section*{Appendix}
\renewcommand{\thesubsection}{A.\Roman{subsection}}
\renewcommand{\thesection}{}
\setcounter{equation}{0}
\numberwithin{equation}{section}
\renewcommand{\theequation}{A.\arabic{equation}}
\setcounter{figure}{0}
\renewcommand{\thefigure}{A.\arabic{figure}}
\renewcommand{\theHfigure}{A.\arabic{figure}}

We have demonstrated the distribution of high-dimensional entanglement through a multi-mode fibre (MMF), exploiting a new method for measuring the transmission matrix of a complex scattering medium with entanglement. In this supplementary material, we provide additional details of our technique and its experimental implementation. To ensure the state coming out of the fibre is entangled in high dimensions, we need to reverse the scrambling effect of the multi-mode fibre on the input entangled state. In Section~\ref{sec:AfterMMF}, we discuss the measurements made on this scrambled state, showing that correlations after propagation through the MMF are lost. 

Our unscrambling technique relies on the concept of channel-state duality, where the information of a channel is mapped onto an output state. Performing phase-stepping measurements on the state after propagation through the MMF, we extract both the amplitude and phase of the complex coefficients of the fibre transmission matrix. Section~\ref{sec:MeasTMDet} gives additional information on this measurement technique and its implementation with spatial light modulators. Using the knowledge of the transmission matrix, we construct and implement a new set of measurements on the state, which reverse the effect of the complex medium and recover correlations in all mutually unbiased bases. This allows us to certify high-dimensional entanglement by lower bounding the fidelity of the output state to an entangled pure target state through the method described in Section~\ref{sec:Fbound}. A detailed summary of the results is given in Section~\ref{sec:results}. 

\begin{figure*}[b]
    \centering
    \includegraphics[width = \textwidth]{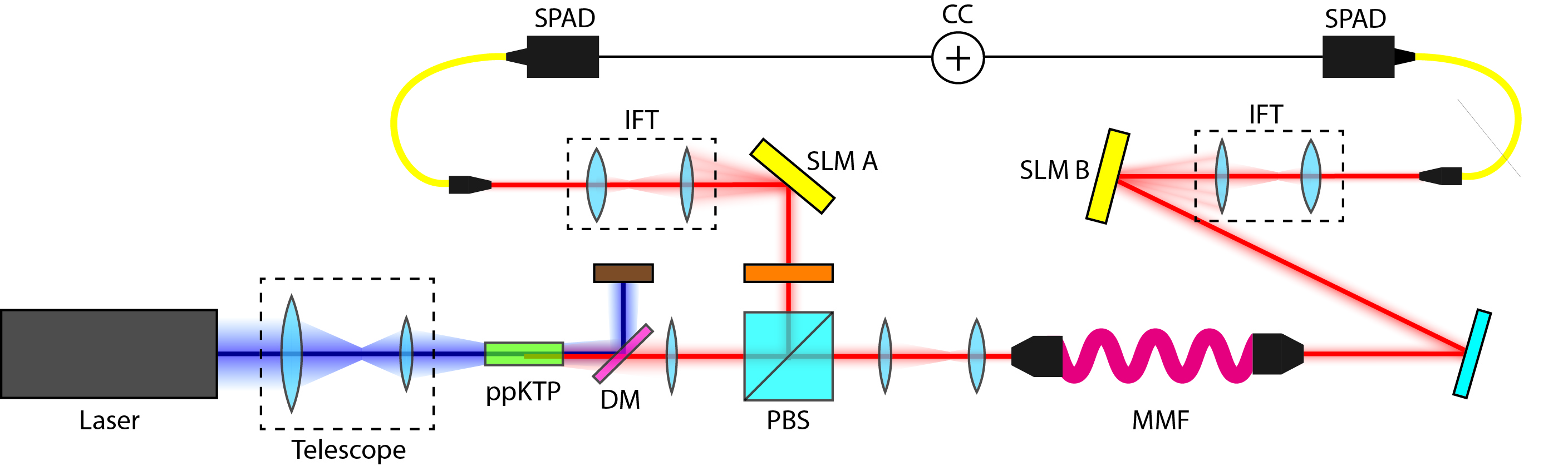}
    \caption{\textbf{Detailed experimental setup}: A 405~nm CW laser is used to pump a 5~mm ppKTP crystal to generate a pair of photons at 810~nm entangled in their transverse position-momentum via Type-II spontaneous-parametric-down-conversion (SPDC). After the pump is removed with a dichroic mirror (DM), the photons are separated by a polarising beam splitter (PBS). The photon corresponding to Alice is made incident on a spatial light modulator (SLM A), where computer generated holograms are displayed in order to manipulate the phase and amplitude of the incident photons. On the other side, the photon corresponding to Bob goes through a multi-mode fibre (MMF) and is then made incident on SLM B. Both photons are demagnified with an intensity-flattening telescope (IFT), coupled to single-mode fibres (SMFs), and detected by single-photon avalanche photodiodes (SPADs). Generalised projective measurements of photonic spatial modes in any chosen Pixel basis are performed via a combination of the SLM, IFT system, SMF, and SPAD. Time-coincident events between the two SPADs are registered by a coincidence counting logic (CC).}
    \label{fig:expSetup}
\end{figure*}

We implement our technique using the experimental setup showed in Fig.~\ref{fig:expSetup}. For measuring our bi-photon state in any given Pixel-mode basis, we use spatial light modulators (SLMs) that allow us to perform generalised projective measurements on the state by coupling selected modes to single-mode fibres. Section~\ref{sec:Normalisation} deals with some of the effects introduced by the use of an SLM, which need to be taken into account for constructing the right unscrambling bases. Finally, in Section \ref{sec:TransportTwoChannels}, we show how our unscrambling method can be extended to the case when each photon of the entangled state is transported through an independent scattering channel.

\subsection{State after the multi-mode fiber}
\label{sec:AfterMMF}

The bi-photon state generated through SPDC is very close to the maximally entangled state. A signature of this entanglement are the two-photon correlations that can be observed in at least two mutually unbiased bases (MUBs), for e.g., the Pixel basis and one of its MUBs. During propagation through an MMF, each spatial mode that enters scatters into all the other modes that the fibre supports. This mixing process results in what we refer to as ``scrambling',' where correlations previously present in these bases are lost.

\begin{figure*}[t!]
    \centering
    \includegraphics[width=\textwidth]{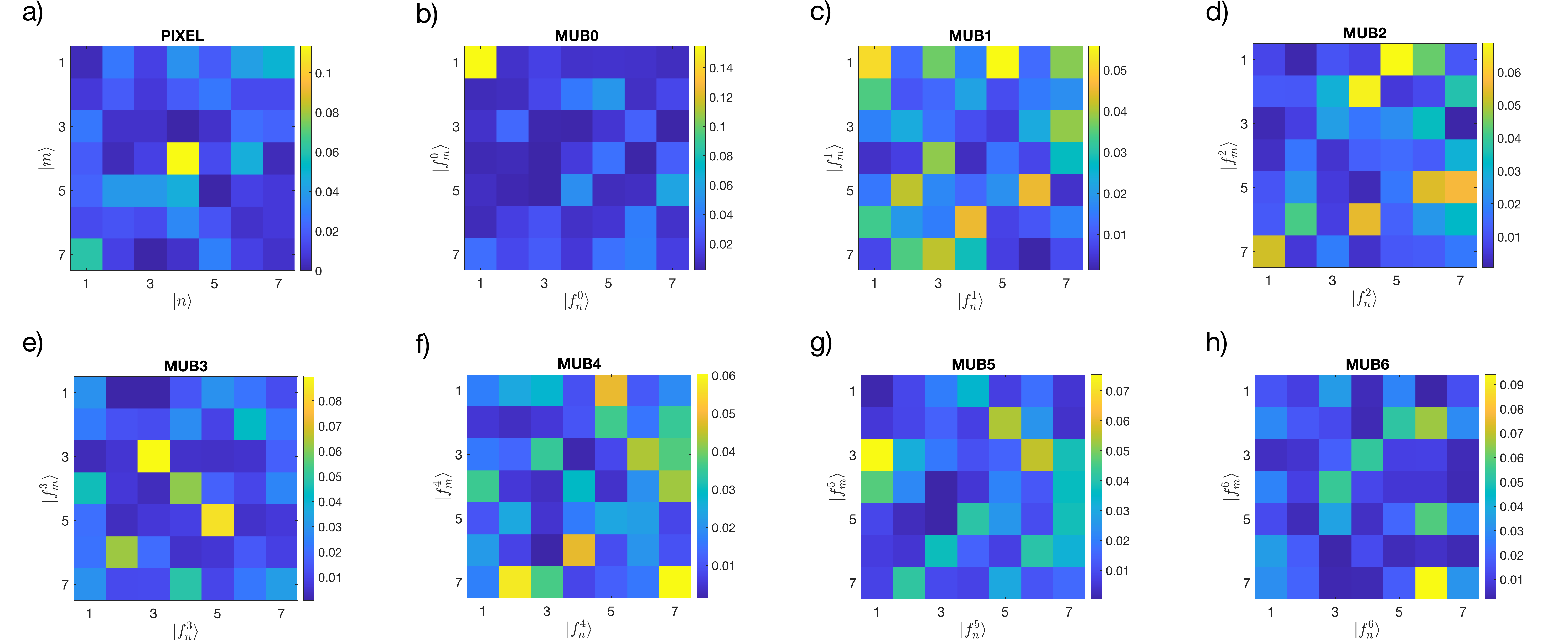}
    \caption{\textbf{``Scrambled'' correlations of the output state $\ket{\psi_{\text{MMF}}}$}: Two-photon coincidence counts in the 7-dimensional Pixel basis  $\{\ket{m},\ket{n}\}_{m,n}$, and its 7 mutually unbiased bases $\{\ket{f_m},\ket{f_n}\}_{m,n}$ at the output of the multi-mode fibre. For this set of measurements we obtained a fidelity to the maximally entangled state of $\tilde{F}(\rho,\psi^+)=5.4 \pm 1.0\%$, i.e., no entanglement can be certified.}
    \label{fig:sMUBs}
\end{figure*}

The effect of the mode-mixing produced by the fibre is depicted in Figure~\ref{fig:sMUBs}. In the absence of the fibre, two-photon correlations measured in the standard Pixel basis and its first MUB result in a fidelity to the seven-dimensional maximally entangled state of $F(\rho,\Phi^{+}) \ge 94.1 \pm 1.2\%$, certifying an entanglement dimensionality of $d_{ent}=7$ (See Fig 2a and 2b in the main text). On the other hand, no entanglement can be certified from measurements made on the state after the multi-mode fibre (Fig.~\ref{fig:sMUBs}). Here, measurements in the Pixel basis ($\{\ket{mn}\}$) and all of its seven MUBs ($\{\ket{f^r_mf^r_n}\}$) are used to obtain the exact fidelity (as opposed to a lower bound) to the maximally entangled state, while also giving the best noise performance. However, these still give us a fidelity of $F(\rho,\Phi^{+})=5.4\pm 1.0\%$, resulting in no entanglement. The details of how to certify entanglement from these measurements are given in the Methods section.

It is clear from the distribution of counts in the correlation matrices of Fig~\ref{fig:sMUBs} that there is a particular mode in which all of the other input modes seem to have scattered, leading to a higher amount of counts in one element of the correlation matrix. This phenomenon arises from the alignment we make before taking the measurement, in which we optimise the efficiency of the coupling by projecting simply the diffraction grating on both SLMs, which leads us to the best results in the experiment. By doing so, we end up optimising the coupling to a particular mode that is close the mode $\ket{4}$ in the Pixel basis (central macropixel), and close to the mode $\ket{f^0_1}$ of the first mutually unbiased bases (all 7 macropixels with the same phase). The effect of coupling to these particular modes is reflected in the measured effective transmission matrix, and will be mapped onto the output state. 

\subsection{Measurement of the transmission matrix}
\label{sec:MeasTMDet}

\begin{figure*}[t!]
    \centering
    \includegraphics[width = \textwidth]{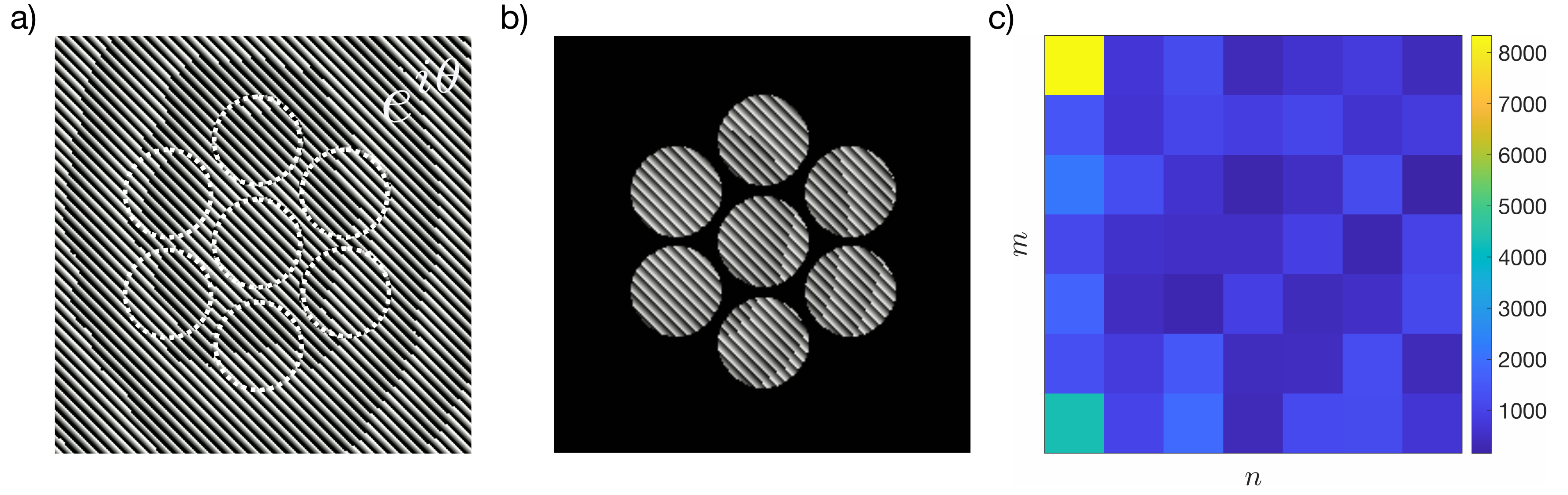}
    \caption{\textbf{$S$ matrix measurement:} Examples of computer-generated holograms displayed on the SLMs corresponding to (a) Alice and (b) Bob, when measuring a given matrix element $R_{mn}^{\theta}$. Changing the relative phase $\theta$, we obtain matrices $R^{\theta}$ that allow us to calculate the elements $S_{mn}$ through Eq. (\ref{eq:Rmn}). The absolute value of the matrix $\mathbf{S}$ is shown in (c).}
    \label{fig:SMeas}
\end{figure*}

Here we show examples of the SLM holograms used in the transmission matrix measurement process, as well as the resulting data. Some equations discussed in the Methods section are repeated for clarity. We measure the transmission matrix $T$ using an internal reference mode that co-propagates and interferes with the macro-pixel modes. By stepping over the relative phase between reference and pixel modes (See Fig.~\ref{fig:SMeas}a and b), we can extract information about the complex amplitudes that the transmission matrix is comprised of. From these measurements we can construct matrices $R(\theta)$ with coefficients of the form:
\be
R_{mn}(\theta)= \left |e^{-i\theta} + t_{nm} \right |^2. 
\label{eq:Rmn}
\ee
Stepping $\theta$ from 0 to $3\pi/2$, we obtain:
\be
S_{mn} = t_{n0} t^*_{nm}= E_n t^*_{nm} 
\ee

\begin{figure*}[t!]
    \centering
    \includegraphics[width = \textwidth]{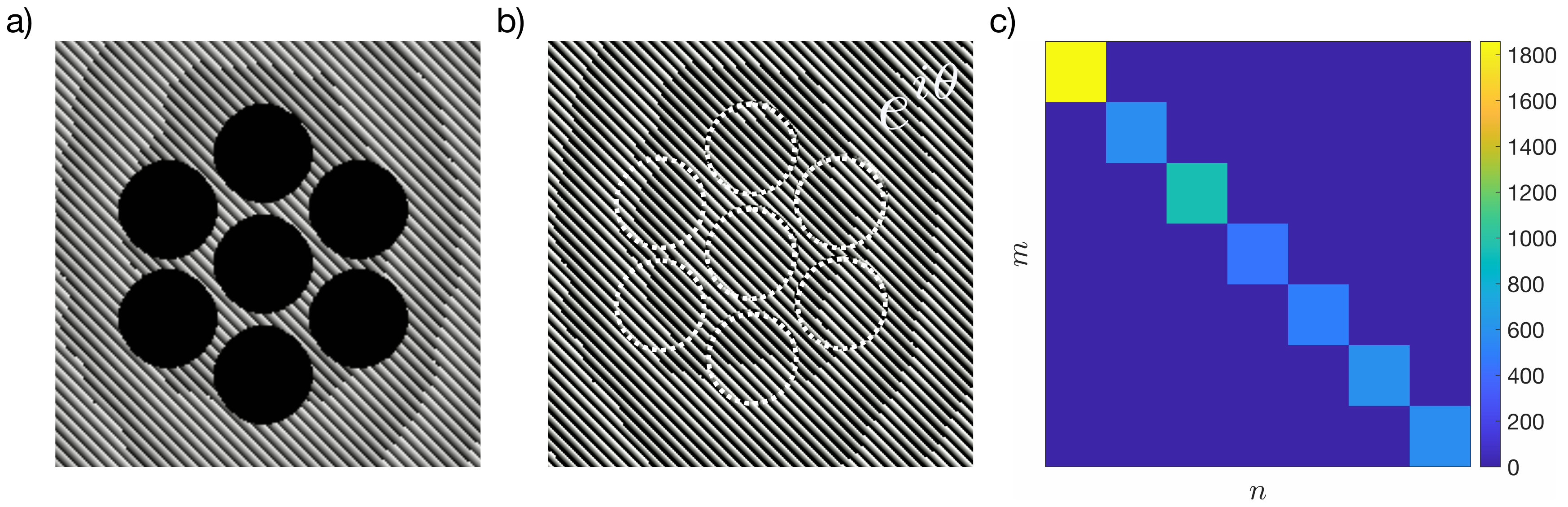}
    \caption{\textbf{$E$ matrix measurement:} Examples of computer-generated holograms displayed on the SLMs corresponding to (a) Alice and (b) Bob, when measuring matrix elements $R_{m}^{\theta}$. We change the relative phase $\theta$ via a phase-stepping process to determine the diagonal elements ($m=n$) of the matrix $\mathbf{E}$. The absolute value of this diagonal matrix is shown in (c).}
    \label{fig:EMeas}
\end{figure*}

\noindent The resulting matrix $S = \{S_{mn}\}$ is related to the actual transmission matrix as follows:
\bea
\mathbf{S} = \mathbf{T^\dagger E},
\label{Smatrix}
\eea 
where $E$ is a diagonal matrix associated with the mixing of the reference and pixel modes after transmission through the MMF. The complex elements of $E$ are determined by displaying the reference mode on Alice, while phase-stepping on Bob (See Fig.~\ref{fig:EMeas}). The results of these measurements are given by:
\be 
R_m^\theta = |t_{00}e^{i\theta}+ t_{m0}|^2. 
\ee
Performing the measurement for different relative phases, we recover a diagonal matrix $\mathbf{E}$ whose absolute value is shown in Fig.~\ref{fig:EMeas}c. With both $S$ and $E$, we can calculate the $T$ matrix characterizing the effect of the MMF as follows:
\be 
\mathbf{T = (SE^{-1})^\dagger }
\label{TMeas}
\ee 

\noindent Using a basis different from the standard basis when performing this phase-stepping technique results in a rotation of the transmission matrix. In our case, we use measurements in the first MUB ($M_0$) in order to maximise our count rates, in which case the obtained transmission matrix is given by:
\be 
T_{M_0} = M_0^* T M_0^T,
\label{eq:Tm}
\ee
where $T$ is the transmission matrix that one would have determined with measurements in the standard basis. The recovered $T_{M_0}$ is shown in Table~\ref{table:tm}.

\begin{smallboxtable}[float = !b]{Measured Transmission Matrix $T_{M_0}$}{tm}
\begin{center}
\begin{tabularx}{\textwidth}{|X|X|X|X|X|X|X|}
\hline
 4.46 - 0.47i &	0.54 + 0.56i &	-0.61 - 0.98i &	0.25 + 0.55i &	-0.29 + 0.90i &	0.62 + 0.30i &	-1.54 + 1.75i \\
\hline
 -1.24 - 0.08i&	0.63 + 0.93i&	-0.01 - 2.16i&	-0.49 - 0.88i&	0.39 + 0.67i&	-1.36 + 0.35i&	-1.35 - 1.17i \\
\hline 
 1.25 - 0.07i&	-0.63 - 0.95i&	0.58 - 0.28i&	-0.50 + 0.23i&	0.18 - 0.21i&	-0.66 + 1.45i&	-0.23 - 2.00i \\
\hline
 0.18 - 0.79i&	1.96 - 0.30i&	-0.57 - 0.01i&	0.45 - 1.08i&	-2.05 - 0.13i&	0.44 - 0.85i&	-0.64 + 0.53i \\
\hline
 -0.52 + 1.13i&	-1.74 - 1.29i&	-0.68 - 0.74i&	-0.17 - 1.89i&	-0.72 - 0.31i&	-0.47 + 0.84i&	-2.40 + 0.09i \\
\hline
 0.56 - 1.24i&	0.95 + 0.46i&	0.52 - 1.96i&	-0.12 - 0.44i&	0.24 - 0.85i&	1.81 - 0.99i&	1.94 + 0.51i \\
\hline
 -0.26 + 0.64i&	1.22 - 0.73i&	0.10 + 0.25i&	-1.05 - 1.37i&	1.80 + 0.80i&	0.53 + 0.34i&	-0.78 - 0.90i
\\
\hline
\end{tabularx}
\end{center}
\small{The complex coefficients of the $7\times 7$ transmission matrix $T_{M_0}$ are shown here. The matrix was determined through a phase-stepping process with measurements in the first mutually unbiased basis of the Pixel basis.}
\end{smallboxtable}

\subsection{Normalisation on the spatial light modulators}
\label{sec:Normalisation}

A hologram for measuring any state in a given basis is calculated as a complex matrix, where each macro-pixel is represented by a complex number. For the appropriate functioning of the SLM, this complex matrix needs to be normalised by the norm of its largest element, which in general is not a problem when using the Pixel, the MUB, or the tilted basis, where all states of one given basis are normalised by the same factor.

When dealing with the new ``scrambling'' bases, adding the transmission matrix to the construction leads to each $i$-th state having a different normalisation factor $\eta_i$. For the states of the standard basis used in Alice, this factor is given by:
\be
\eta_i = \text{max}_i\left|(T_{M_0}^{-1})^T M_0 \right|_{ij},
\ee
where we take the largest norm of each row of the unscrambling operator applied on Alice.\\ 

\noindent The normalisation made for the functioning of the SLM modifies the unscrambling operator for the standard basis as:
\be
\mathbf{W^{\eta}_{M}} = \left(\eta^{-1}(T_{M_0}^{-1})^TM_0  \otimes M_0^*\right),
\ee 
where the matrix $\eta$ is a diagonal matrix composed of the $\eta_i$ factors. With this modified operator, we have an unscrambled state given by:
\bea
\ket{\psi}^{\eta}_{\text{u}} &=& \mathbf{W^{\eta}_M}\ket{\psi_{\text{MMF}}}  \nonumber \\
&=& \left(\eta^{-1}(T_{M_0}^{-1})^TM_0  \otimes M_0^*\right) \ket{\psi_{\text{MMF}}} \nonumber \\
&=& \eta^{-1}\ket{\Psi^+}.
\eea
Measurements of this unscrambled state lead to a ``non-flat'' distribution of the correlated counts that is dictated by the normalisation factors. To perform mutually unbiased measurements on this same state, we also modify the operator $\mathbf{V^{\eta}_M}$ to:
\bea 
\mathbf{V^{\eta}_M} &=& \mathbf{M_r W^{\eta}_M} \nonumber \\
&=& (M_r\otimes M_r^*)\left(\eta^{-1}(T_{M_0}^{-1})^T M_0  \otimes M_0^*\right).
\eea
Once again, this operator defines a new basis to be used at Alice, where each state is projected by the SLM with a second normalisation factor:
\be
\zeta_i = \text{max}_i\left|M_r\eta^{-1}(T_{M_0}^{-1})^T M_0 \right|_{ij}.
\ee
 The latter normalisation leads to obtaining two-photon coincidences for the MUB basis given by:
\bea
\tilde{N}^{\zeta}_{wv} 
&=& |\zeta^{-1}\braket{g^{\zeta}_wg^{\zeta}_v}{\psi_{\text{MMF}}}|^2 \nonumber \\
&=& |\zeta^{-1}\brakket{mn}{\mathbf{V^{\eta}_M}}{\psi_{\text{MMF}}}|^2 \nonumber \\
&=& |\zeta^{-1}\brakket{mn}{\mathbf{\mathbf{M_r W^{\eta}_M}}}{\psi_{\text{MMF}}}|^2 \nonumber \\
&=& |\zeta^{-1}\braket{f_mf_n}{\psi_{\text{u}}^{\eta}}|^2 \nonumber \\ &=&|\zeta^{-1}|^2|\braket{f_mf_n}{\psi_{\text{u}}^{\eta}}|^2.
\eea 
The normalisation made for the functioning of the SLMs cannot be avoided. The factor of $\eta$ will affect the Schmidt coefficients of the target state that are chosen from measurements in the standard basis, and hence, it will contribute to how the unscrambled state looks like. For certifying entanglement in this unscrambled state, is important to take the normalisation of $\eta$ into account when defining the operator for unscrambling in the mutually unbiased bases. However, the second measurement has a factor $\zeta$ that has nothing to do with the state, and is thus corrected in post-processing.

\subsection*{Certification of entanglement}
\label{sec:Fbound}

In order to certify the presence of high-dimensional entanglement in the recovered state after the multi-mode fibre, we employ a recently developed witness that uses correlations in at least two mutually unbiased bases~\cite{Bavaresco:2018gw}. One can determine a lower bound for the fidelity $F(\rho,\Phi)$ of a given state $\rho$ to a pure bipartite target state $\ket{\Phi}$, using measurements in only two bases. Since the fidelity to a target entangled state also provides information about the dimensionality of entanglement, we use this bound for certifying the entanglement dimensionality of the state at the output of the fibre.

Consider a target state that is written in its Schmidt or standard basis as:
\be
\ket{\Phi} = \sum_{m=0}^6 \lambda_m \ket{m}_A\ket{m}_B,
\ee
where we can either nominate a target state using the Schmidt coefficients obtained from two-photon correlations measured in the standard basis:
\be
\lambda_m = \sqrt{\frac{\brakket{mm}{\rho}{mm}}{\sum_n \brakket{nn}{\rho}{nn}}},
\label{eq:SchmidtC}
\ee
or consider the maximally-entangled state $\ket{\Psi^{+}}$ as a target state, which has $\lambda_i = 1/\sqrt{d}$ for every $i$.

The choice of the target state will determine which second basis should be used for calculating the lower bound. If the target state is chosen to be maximally entangled, one performs measurements in a mutually unbiased basis (MUB). On the other hand, if the target state is chosen to be a non-maximally entangled state (estimated from measurements in the standard basis), one uses the tilted bases that are constructed using the $\lambda_m$ coefficients as described in the Methods. 

For any state $\rho$ with Schmidt rank $k\le d$, the fidelity to a target state $F(\rho,\Phi)$ is bounded by:
\be
F(\rho,\Phi) \le B_k(\Phi) := \sum_{m=0}^{k-1}\lambda_{i_m}^2,
\ee
where the sum runs over the $k$ largest Schmidt coefficients of $\Phi$, that is, $i_m$, with $m=0,...,d-1$ such that $\lambda_{i_m} \ge \lambda_{i_m'} \forall m \le m'$. The experimental fidelity bound $\tilde{F}(\rho,\Phi)$ that can be determined with measurements in two bases fulfills the following dimensionality witness inequality:
\be
\tilde{F}(\rho,\Phi) \le F(\rho,\Phi) \le B_k(\Psi).
\ee
For unscrambling the entanglement after the state goes through the MMF, we have constructed a new standard basis $\{\ket{wv}\}_{w,v=0,1,...,d-1}$ by applying the operator $\mathbf{W_M}$ to the pixel modes. 
Measuring the $\ket{w}$ state at Alice and the $\ket{v}$ state in Bob results in two-photon coincidence counts $\mathcal{N}_{wv}$ 
that allow us to calculate the diagonal elements of the density matrix as:
\be 
\brakket{wv}{\rho}{wv} = \frac{\mathcal{N}_{wv}}{\sum_{k,l}\mathcal{N}_{kl}}.
\label{eq:rhoCOMP}
\ee 
The cross-talk matrix containing these elements display correlations in the Pixel basis. From these measurement, we can use Eq. \ref{eq:SchmidtC} to calculate the probability amplitudes of the target state, i.e., the nominated Schmidt coefficients.

On the other hand, using the new mutually unbiased bases $\{\ket{g^r_wg^r_v}\}_{w,v=0,1,...,d-1}$ indexed by $r=0,...,d-1$, we obtain two-photon coincidences counts $\mathcal{\tilde{N}}_{wv}$ 
from which we construct the density matrix elements given by:
\be
\brakket{g^r_w g^{r*}_v}{\rho}{g^r_w g^{r*}_v} = \frac{\mathcal{\tilde{N}}_{wv}}{\sum_{k,l}\mathcal{\tilde{N}}_{kl}}
\label{eq:rhoMUB}
\ee 
For each chosen basis $r$, these density matrix elements can be put in the form of cross-talk matrices that display correlations in the corresponding MUB.

As mentioned above, the choice of the Schmidt coefficients $\lambda_m$ determines whether we want the target state to be $\ket{\Phi}$ or $\ket{\Psi^+}$. In our measurement, this is reflected by how we construct the operator $\mathbf{V_M}$. For a target state $\ket{\Phi}$, measurements are made with the new tilted bases $\{\ket{\tilde{g}^r_w\tilde{g}^r_v}\}_{w,v=0,1,...,d-1}$, which are constructed by having the tilted basis operator $\mathbf{\tilde{M}_r}$, instead of the MUB basis operator $\mathbf{M_r}$. 
In this case, the obtained density matrix elements are constructed from the coincidence counts as follows:
\be
\brakket{\tilde{g}^r_w \tilde{g}^{r*}_v}{\rho}{\tilde{g}^r_w \tilde{g}^{r*}_v} = c_{\lambda}\frac{\mathcal{\tilde{N}}_{wv}}{\sum_{k,l}\mathcal{\tilde{N}}_{kl}}
\label{eq:rhoTILT}
\ee 

The factor $c_{\lambda}:=\frac{d^2}{(\sum_{k}\lambda_k)^2}\sum_{m,n}\lambda_m\lambda_n\brakket{mn}{\rho}{mn}$ here is used for normalisation that is required because of the non-orthogonality of the tilted bases \cite{Bavaresco:2018gw}.

With the density matrix elements determined from Eq.~\ref{eq:rhoCOMP} and either Eq.~\ref{eq:rhoMUB} or \ref{eq:rhoTILT}, we can calculate the fidelity bound proposed in Ref.~\cite{Bavaresco:2018gw}, and certify the presence of high-dimensional entanglement by only performing measurements after the MMF. Furthermore, when all $d$ MUBs, or all $d$ tilted bases are used, the fidelity bound becomes tight, that is, we obtain an estimate for the exact fidelity: $\tilde{F}(\rho,\Phi) = F(\rho,\Phi)$.

\subsection{Unscrambled data}
\label{sec:results}

\begin{figure*}[t!]
    \centering
    \includegraphics[width = \textwidth]{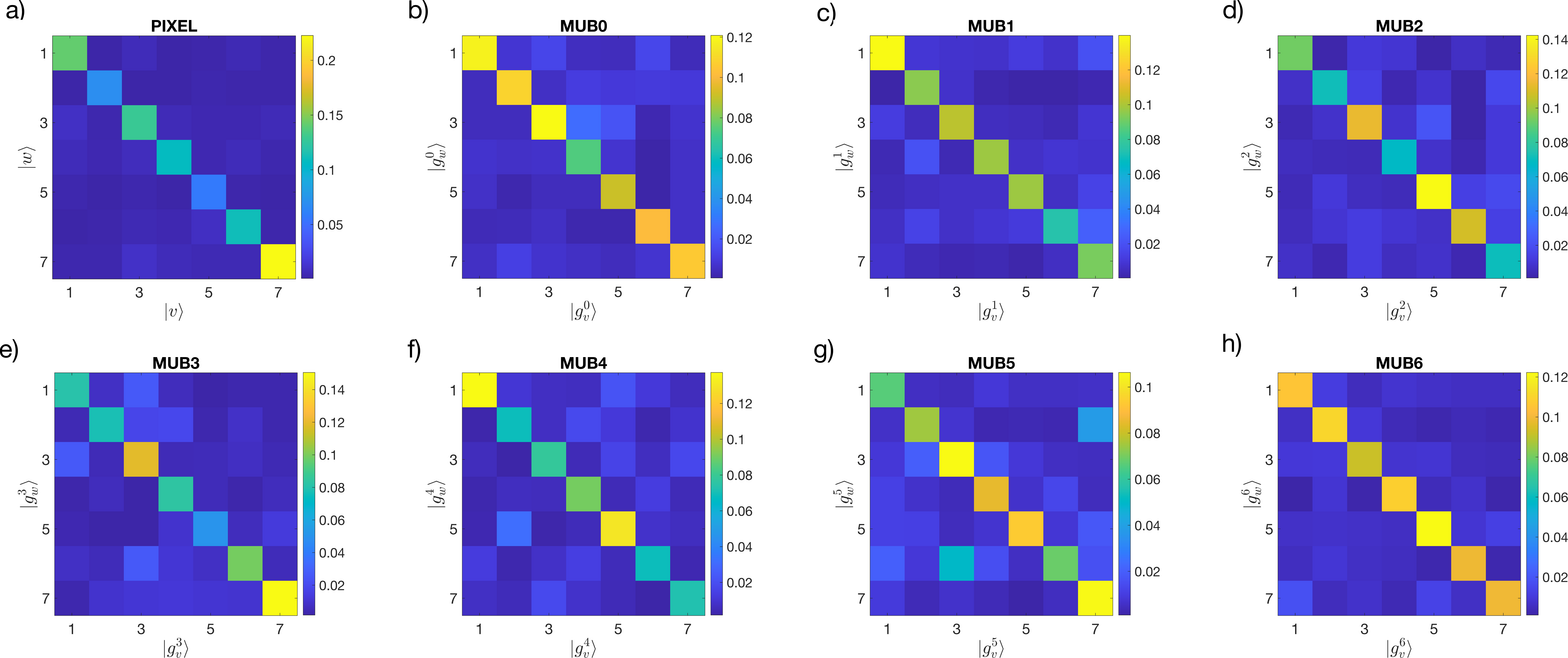}
    \caption{\textbf{Experimental data certifying 5-dimensional entanglement through measurements in MUBs}: Normalised two-photon coincidence counts showing recovered correlations in the 7-dimensional Pixel basis $\{\ket{w},\ket{v}\}_{w,v}$ and its 7 mutually unbiased bases $\{\ket{g_w},\ket{g_v}\}_{w,v}$ after propagation through the multi-mode fibre. From these sets of measurements, we calculate a fidelity to the $d=7$ maximally entangled state of $\tilde{F}(\rho,\Phi^{+})=65.3 \pm 0.8\%$ , yielding an entanglement dimensionality of $d_{\text{ent}}=5$}
    \label{fig:MUBs}
\end{figure*} 

Using the dimensionality witness described in the previous section, measurements in mutually unbiased bases allow us to certify the high-dimensional entanglement that has been transported through the fibre. In the Methods, we showed how the transmission matrix allows us to construct a new set of bases for measuring the state at the output of the fibre. Since the scattering information from the fibre is encoded in these new bases, these measurements invert the scattering process inside the complex medium and result in correlations that are unscrambled.

\begin{smallboxtable}[float = !b]{Measured $\lambda_{m}$ Values}{lambdas}
\begin{center}
\begin{tabularx}{\textwidth}{|X|X|X|X|X|X|X|}
\hline
$\lambda_0$ 	& $\lambda_1$ & $\lambda_2$	& $\lambda_3$ & $\lambda_4$ & $\lambda_5$ 	& $\lambda_6$ \\
\hline
0.4079 	& 	0.2930 		& 	0.3118		&	0.3553	  & 0.3596 	 	& 	0.4329 		& 	0.4556		\\
\hline
\end{tabularx}
\end{center}
\small{Measured probability amplitudes used for nominating the target state $\ket{\Phi}=\sum_{m=0}^{6}\lambda_{m}\ket{mm}$ and subsequently constructing the titled bases. Values of $m=0...6$ label the corresponding pixel mode.}
\end{smallboxtable}

\begin{figure*}[t!]
    \centering
    \includegraphics[width=\textwidth]{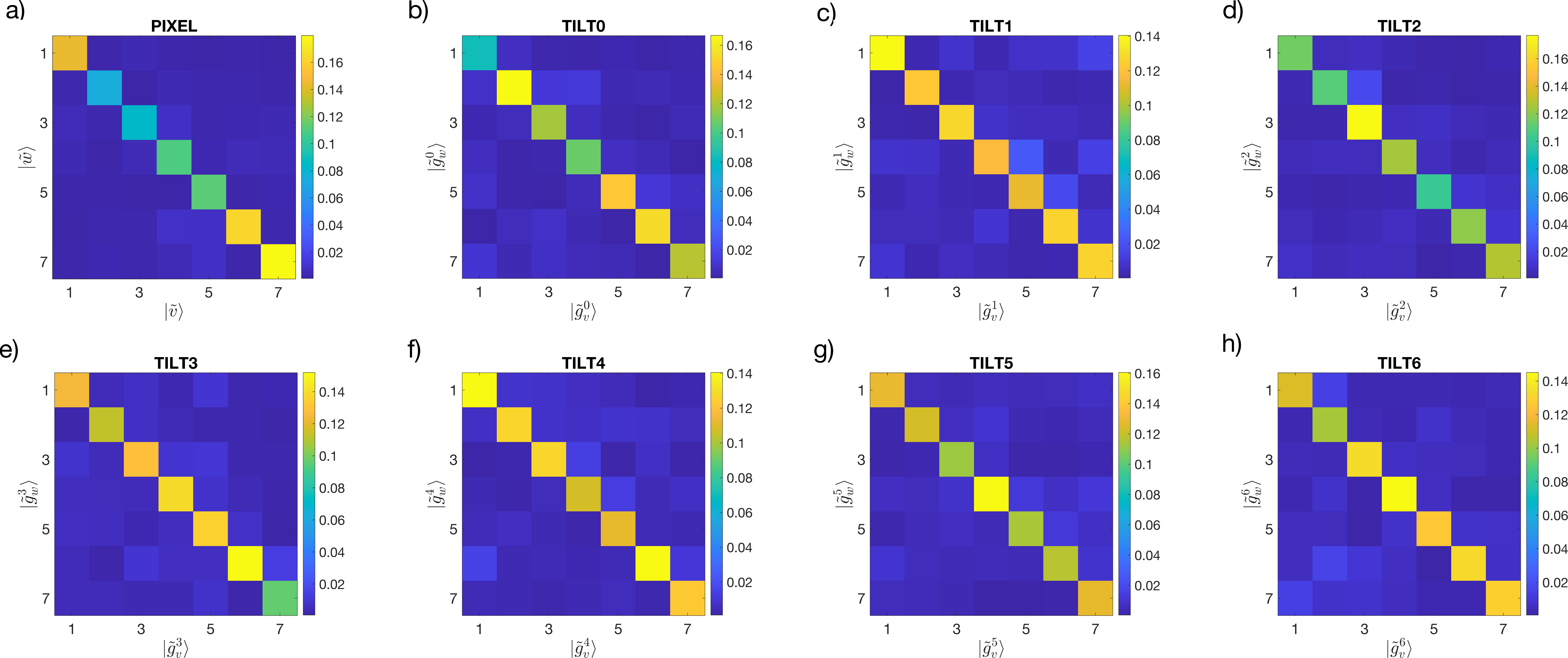}
    \caption{\textbf{Experimental data certifying 6-dimensional entanglement through measurements in tilted bases}: Normalised two-photon coincidence counts showing recovered correlations in the 7-dimensional Pixel basis $\{\ket{{w}},\ket{v}\}_{w,v}$ and 7 MUB-like tilted bases $\{\ket{\tilde{g}_w},\ket{\tilde{g}_v}\}_{w,v}$ after propagation through the multi-mode fibre. From these sets of measurements we calculate a fidelity to the $d=7$ target state of $\tilde{F}(\rho,\Phi)=84.4 \pm 1.8\%$ , yielding an entanglement dimensionality of $d_{\text{ent}}=6$}
    \label{fig:TILTs}
\end{figure*}

In Fig.~\ref{fig:MUBs} we can see the unscrambled two-photon correlation matrices in all $d+1=8$ mutually unbiased bases. From these measurements, we obtain a fidelity to the 7-dimensional maximally entangled state of $F(\rho,\Phi)=65.3 \pm 0.8\%$. 
Taking into account that $B_4(\Phi^*)=4/7=0.5714$, these data allow us to certify an entanglement dimensionality of $d_{ent}=5$. Figure \ref{fig:TILTs} shows unscrambled two-photon correlation matrices for the standard and the 7 tilted bases. As discussed in the Methods, these tilted bases are constructed from Schmidt coefficients obtained from correlations measured in the standard basis (See Table~\ref{table:lambdas}). From these measurements, we obtain a fidelity to the target state of $F(\rho,\Phi)=84.4 \pm 1.8\%$, 
which violates the bound $B_5(\Phi)=\sum^{4}_{i=0} \lambda_i^2  = 0.8169$. This fidelity thus certifies an entanglement dimensionality of $d_{ent}=6$.

\subsection{Transporting each photon through an independent channel}
\label{sec:TransportTwoChannels}
Consider the case in which both photons of the generated pair are transported through a separate MMF channel. In this case, we will have two different unitary transformations, each representing the scattering produced by one of the two fibres. One unitary operator operates on Alice in the following manner:
\be
\hat{U}_A = \sum_{k'r'}U_{k'r'}\ket{k'}\bra{r'},
\ee 
where $\ket{k'}_{A}$, and $\ket{r'}_{A}$ are elements of a basis in Alice's subspace. The other unitary operates on Bob as follows:
\be
\hat{U}_B = \sum_{kr}U_{kr}\ket{k}\bra{r},
\ee 
where $\ket{k}_{B}$, and $\ket{r}_{B}$ are elements of a basis in Bob's subspace.

For simplicity, we take the state before the fibre to be the maximally entangled state. After the fibre, we have a state given by:
\bea 
\ket{\psi_{\text{MMF}}} &=& (\hat{U}_A \otimes \hat{U}_B)\ket{\Psi^+} = (\hat{I} \otimes \hat{U}_B(\hat{U}_A)^T)\ket{\Psi^+} \nonumber \\
&=& (I \otimes \hat{T})\ket{\Psi^+},
\label{eq:output2MMF}
\eea
where using the property of applying operations on a maximally entangled state as a one-sided operation on either of its parts, the case of two different fibres can be reduced to a single effective transmission matrix or channel given by
\be
\hat{T} = \hat{U}_B(\hat{U}_A)^T,
\ee
which operates only on one of the photons of the entangled pair. Taking into account Eq.~\ref{eq:output2MMF}, the information of both $\hat{U}_A$ and $\hat{U}_B$ is encoded into the effective transmission matrix $\hat{T}$, which has been mapped onto the output state \ket{\psi_{\text{MMF}}}, which is analogous to the one obtained in our experiment. 

The procedure outlined above allows us to determine this transmission matrix, reverse its effect by only performing local operations on one photon, and successfully transport high-dimensional entanglement through an MMF. The extension of our methods to the case discussed above of entanglement transmission through two independent complex scattering channels, where the effects of both channels can be inverted with operations on only one photon is of fundamental interest and could hold significance for quantum communication protocols where one may not have access to both photons after transmission.

\end{document}